\documentclass[twocolumn,showpacs,preprintnumbers,amsmath,amssymb]{revtex4}
\usepackage{epsfig}
\usepackage{amsmath}

\begin{document}

\title{Avalanches of Activation and Spikes in Neuronal Complex Networks}

\author{Luciano da Fontoura Costa}
\affiliation{Institute of Physics at S\~ao Carlos, University of
S\~ao Paulo, PO Box 369, S\~ao Carlos, S\~ao Paulo, 13560-970 Brazil}

\date{31st Jan 2008}

\begin{abstract}
As shown recently (arXiv:0801.3056), several types of neuronal complex
networks involving non-linear integration-and-fire dynamics exhibit an
abrupt activation along their transient regime.  Interestingly, such
an avalanche of activation has also been found to depend strongly on
the topology of the networks: while the Erd\H{o}s-R\'eny,
Barab\'asi-Albert, path-regular and path-transformed BA models exhibit
well-defined avalanches; Watts-Strogatz and geographical structures
present instead a gradual dispersion of activation amongst their
nodes.  The current work investigates such phenomena by considering a
mean-field equivalent model of a network which is strongly founded on
the concepts of concentric neighborhoods and degrees.  It is shown
that the hierarchical number of nodes and hierarchical degrees define
the intensity and timing of the avalanches.  This approach also
allowed the identification of the beginning activation times during
the transient dynamics, which is particularly important for community
identification (arXiv:0801.4269, arXiv:0801.4684).  The main concepts
and results in this work are illustrated with respect to theoretical
and real-world (\emph{C. elegans}) networks.  Several results are
reported, including the identification of secondary avalanches, the
validation of the equivalent model, the identification of the possible
universality of the avalanches for most networks (depending only on
the network size), as well as the identification of the fact that
different avalanches can be obtained by locating the activation source
at different neurons of the \emph{C. elegans} network.
\end{abstract}

\pacs{87.18.Sn, 05.40Fb, 89.70.Hj, 89.75.Hc, 89.75.Kd}
\maketitle

\vspace{0.5cm}
\emph{`I have also thought of a model city from which I deduce all
the others.' 
(Invisible Cities, I. Calvino)}

\section{Introduction}

Investigations addressing the relationship between structure and
dynamics have become increasingly important in several scientific
areas, from neuroscience (e.g.~\cite{Ascoli:1999,Costa_revneur:2005})
to complex networks (e.g.~\cite{Newman:2003, Boccaletti:2006}), and
more recently to both these areas (e.g.~\cite{Stauffer_Hopfield,
Stauffer_Costa, Costa_revneur:2005, Kim:2004, Timme:2006, Osipov:2007,
Hasegawa:2004, Hasegawa:2005, Park:2006}).  While a great deal of
attention has been recently focused on investigations or relationships
between linear synchronization and network topology
(e.g.~\cite{Boccaletti:2006,Hong_etal:2004,Lee:2005, Zhou:2006,
Boccaletti:2007, Takashi:2006, Sorrentino:2007, Almendral:2007}),
relatively lesser attention has been directed towards studying the
structure-dynamics paradigm by considering non-linear and/or transient
regimes.  In a recent series of articles~\cite{Costa_nrn:2008,
Costa_begin:2008,Costa_activ:2008}, the transient dynamics of complex
neuronal networks composed of simple integrate-and-fire neurons has
been shown to yield a series of remarkable effects, including
avalanches of activation~\cite{Costa_nrn:2008} and activation
confinement inside topological communities~\cite{Costa_begin:2008,
Costa_activ:2008}.

\begin{figure*}[htb]
  \vspace{0.3cm} 
  \begin{center}
  \includegraphics[width=1\linewidth]{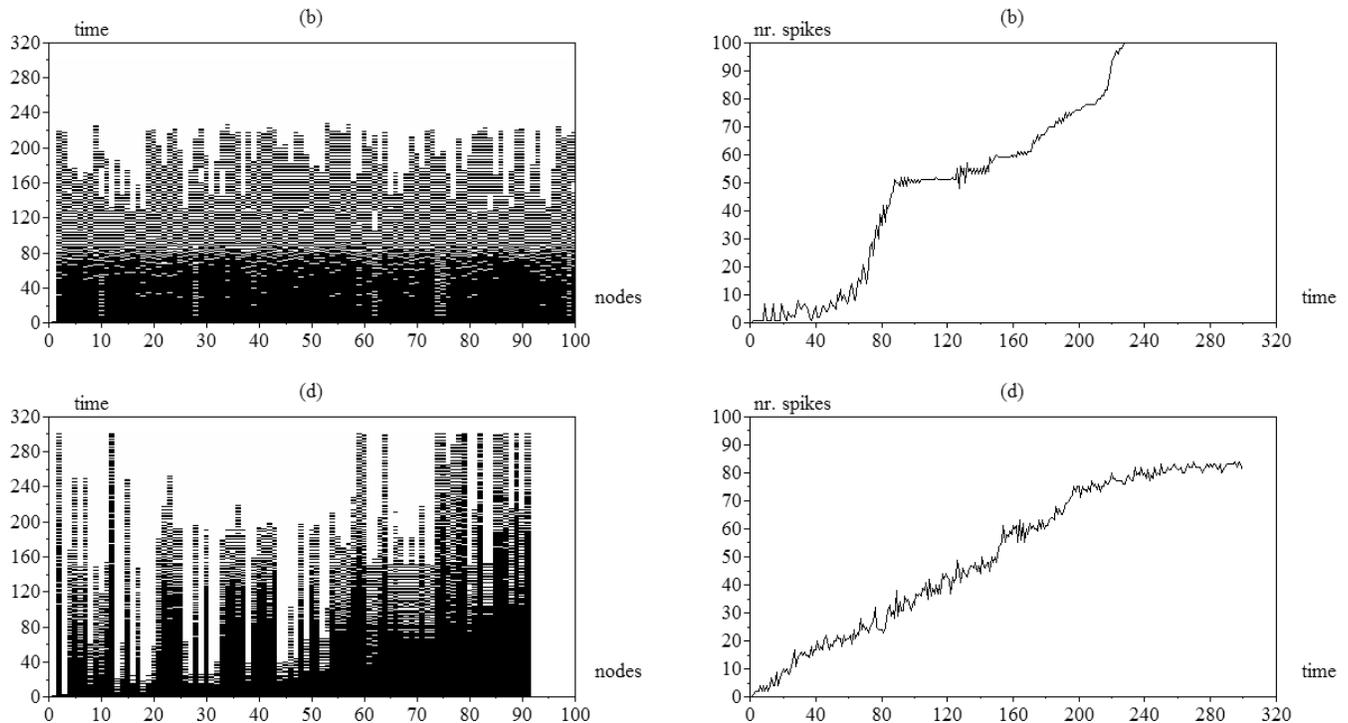} 
  \caption{The spikes at each neuron along time (spikegram) for 
              a path-regular (a)
              and a geographical (c) network.  The respective total 
              number of spikes along each time step during the
              transient activation (300 initial steps) are shown in 
              (b) and (d), respectively. The path-regular network 
              contained 100 nodes and the geographical 91 nodes.
              Both networks had average degree equal to 6. While two
              avalanches are observed for the PN network, the GG 
              dynamics involves only a gradual and smooth increase of spiking.
  }~\label{fig:ex_avals} \end{center}
\end{figure*}

As external activation is fed into the network through a selected
node, which acts as a source, it spreads through the network in a way
which is highly dependent on the network topology.  While
Watts-Strogatz (WS) and geographical (GG) networks are characterized
by more progressive, gradual dispersion of the activation from the
source to the other nodes, Erd\H{o}s-R\'enyi (ER), Barab\'asi-Albert
(BA), path-regular~\footnote{This type of network exhibits a
particularly regular structure (almost identical degree for each
node), being composed of paths proceeding through all network nodes,
without repetition~\cite{Costa_comp:2007}.}  (PN) and path-transformed
BA networks (PA) structures yield an abrupt onset of activation after
some initial time steps, after which the whole network becomes
intensely activated~\cite{Costa_nrn:2008}.  Figure~\ref{fig:ex_avals}
illustrates this phenomenon with respect to the total number of spikes
along time obtained for the transient activation period (300 initial
steps) in a PN (Fig.~\ref{fig:ex_avals}a-b) and a GG
(Fig.~\ref{fig:ex_avals}c-d) with similar sizes and average degrees.
Observe the presence of a secondary avalanche taking place at later
times for the PN case (Fig.~\ref{fig:ex_avals}a-b).

In addition, the times for each neuron to receive the first non-zero
activation, through synapses, were found to be particularly related to
the community organization of networks~\cite{Costa_begin:2008}.  These
findings have several implications for biological and computational
neuroscience and community identification, including the fundamental
role that the integration-and-fire dynamics have for confinement of
activation within neuronal modules, avoiding widespread activation of
whole neuronal areas.  The important point here concerns the
simultaneous localization of neuronal activity along time and space
implied by the non-linearities.  However, these results represent
particularly interesting subjects also for more basic investigations.

The present work reports a combined theoretical and experimental
analysis of the activation dynamics in complex networks of varying
types by considering two fundamental concepts.  First, the
\emph{concentric} (or hierarchical~\footnote{Though the name
\emph{hierarchical} was initially (e.g.~\cite{Costa:2004, Costa_NJP:2007, 
Costa_JSP:2006, Costa_EPJB:2005}) used to express the successive
neighborhoods around each node of a complex network, the term
\emph{concentric} was later also adopted in order to distinguish from 
hierarchical networks (e.g.\cite{Ravasz:2003}).  Both these terms are
used in this work.})  neighborhoods around the source node are
identified and characterized in terms of the respective number of
nodes, hierarchical degree and intra-ring degree~\cite{Costa:2004,
Costa_NJP:2007, Costa_JSP:2006, Costa_EPJB:2005}.  Such a concentric
representation of the original network is fundamental for the study of
complex neuronal network dynamics because the activation in such
non-linear systems progress precisely along the concentric
neighborhoods.  The concentric representation of the original network
is then used to derive an \emph{equivalent} mean-field model of the
original network, namely a chain complex neuronal network.  It is
shown analytically that the avalanches of activation along the
transient regime of the integrate-and-fire neuronal networks is an
immediate consequence of the number of nodes at each concentric
level.

This article starts by briefly reviewing the main complex neuronal
network concepts, the adopted 7 theoretical models of complex
networks, as well as the concentric representation of networks and
respective hierarchical measurements.  It proceeds by developing the
mean-field equivalent model and by showing the intrinsic relationships
between the hierarchical structure of networks and the properties of
the avalanches.  This approach also allowed important results
regarding the characterization of the beginning activation times,
which have been found to be intrinsically related to community
structure~\cite{Costa_begin:2008, Costa_activ:2008}.

\section{Basic Concepts}

A undirected complex network with $N$ nodes can be represented in
terms of its respective \emph{adjacency matrix} $K$, such that the
presence of an edge between nodes $i$ and $j$ is represented as
$K(i,j)=K(j,i)=1$, with $K(i,j)=K(j,i)=0$ being otherwise imposed.
The \emph{immediate neighbors} of a node $i$ are those nodes which
share an edge with $i$, i.e. are at distance 1 from $i$.  The
\emph{degree} of a node $i$, henceforth represented as $k(i)$, is equal 
to the number of its immediate neighbors. In a directed network, the
\emph{in-degree} of a node is defined as being equal to the number of 
incoming edges, while the out-degree corresponds to the number of
outgoing edges.  Two nodes are said to be
\emph{adjacent} if they share an edge.  Two edges are adjacent if they
share a node.  A \emph{walk} is any linear sequence of adjacent edges.
A \emph{path} is a walk which never repeats any node or edge.  The
length of a walk (or path) corresponds to its respective number of
edges.    

The \emph{h-neighborhood} of a node $i$ is the set of nodes which are
at shortest path distance of $h$ edges from node $i$.  Such successive
neighborhoods define the concentric (or hierarchical) levels $h$ of
the network with respect to the reference node $i$.  The number of
nodes in such a neighborhood is henceforth represented as
$n_h(i)$. Each of such nodes is an \emph{h-neighbor} of $i$.  The
\emph{hierarchical degree}~\cite{Costa:2004} of a node $i$, henceforth
expressed as $k_h(i)$, is the number of edges between its
$h-$neighbors and its $(h+1)-$neighbors.  The \emph{h-th intra-ring
degree}~\cite{Costa:2004,Costa_NJP:2007, Costa_JSP:2006,
Costa_EPJB:2005} of a node $i$, denoted here as $a_h(i)$, is the
number of edges among the nodes of the $h-$neighbors of $i$.

\begin{figure}[htb]
  \vspace{0.3cm} 
  \begin{center}
  \includegraphics[width=0.9\linewidth]{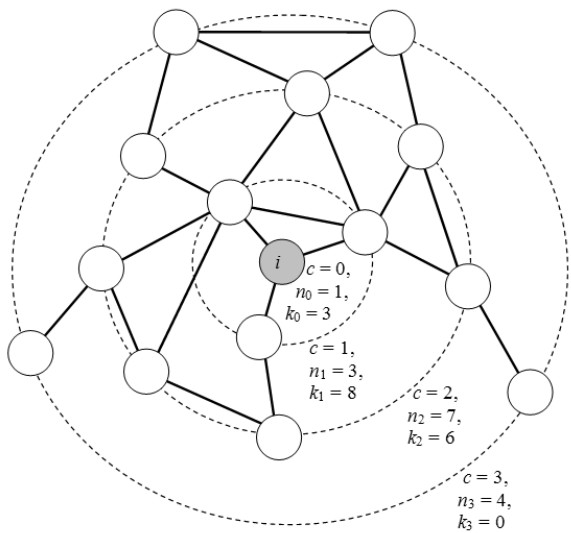} 
  \caption{The concentric organization of a simple network
             with respect to the reference node $i$.  The hierarchical
             number of nodes $n_h(i)$, hierarchical degrees $k_h(i)$
             and intra-ring degrees $a_h(i)$ provide a rich 
             characterization of the connectivity around the reference
             node.
  }~\label{fig:ex_conc} 
  \end{center}
\end{figure}

Figure~\ref{fig:ex_conc} illustrates the concentric organization of a
simple network with respect to the reference node $i$.  Observe that
different concentric organizations will be obtained for the other
nodes.  The organization obtained for the reference node contains 3
concentric levels (also called rings) defined by 3 successive
neighborhoods, respectively with $n_0(i)=1$, $n_1(i)=3$, $n_2(i)=7$,
and $n_3(i)=4$ nodes.  The hierarchical degree of node $i$ at the
level $c=0$ is identical to its traditional degree.  However, now we
have additional information about the connectivity around the
reference node provided by its hierarchical degrees $k_1(i) = 8$;
$k_2(i) = 7$ and $k_c(i) = 0$ and \emph{k-th intra-ring degrees}
$a_0(i)=0$; $a_1(i)=1$; $a_2(i)=3$; and $a_3(i)=1$.

We consider 7 theoretical models of complex networks corresponding to
the Erd\H{o}s-R\'eny (ER), Barab\'asi-Albert (BA), Watts-Strogatz
(WS), geographical (GG), path-regular (PN), path-transformed BA (PA)
as well as version of the PN model, henceforth abbreviated as $PI$,
where all nodes have \emph{identical} degrees.  The ER networks are
obtained by implementing edges between pairs of nodes with fixed
probability.  The BA network starts with $m0$ nodes and additional
nodes are attached with $m$ edges each, which are attached to the
previous nodes with probability proportional to their respective
degrees (~\cite{Albert_Barab:2002}).  The WS structures are obtained
by by rewiring $10\%$ of the edges in a linear regular lattice with
suitable degree.  The PN and PA models belong to the family of
\emph{knitted networks}~\cite{Costa_comp:2007}.  The path-regular
networks (PN) are constructed by performing paths through all nodes in
the network, without repetition.  The path-transformed BA model (PA)
is obtained through the star-path
transformation~\cite{Costa_path:2007} of a BA network with the same
number of nodes and similar average degree.  The PI mode is introduced
in this article as a version of the PN model where all nodes have
\emph{identical} degrees.  This is obtained through the two following
modifications of the PN algorithm: (a) one path can not go through an
edge visited by a previous path; and (b) the initial and final nodes
of each path are connected.  This new model is particularly suitable
for illustrating the developments reported in this article.  All used
networks have similar number of nodes and edges, and only the nodes
corresponding to the largest connected component are taken into
account.  However, because of the relatively large average node
degrees adopted henceforth, most nodes are generally included in the
largest components.

\begin{figure}[htb] 
  \begin{center}
  \includegraphics[width=0.9\linewidth]{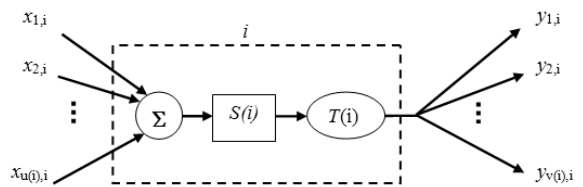} 
  \caption{The integrate-and-fire model of neuron $i$ adopted in this
              work incorporates three main elements: integrator $\Sigma$,
              activation state $S(i)$ and threshold $T(i)$.  Observe 
              that the number of dendrites $u(i)$ and axons $v(i)$ 
              are equal to the in-and out-degree of node $i$, respectively.
  }~\label{fig:neuron} 
  \end{center}
\end{figure}

A \emph{complex neuronal network} is a neuronal network whose
connectivity is given by a complex neuronal network.  Each neuron is
represented as a node, and each synapse as an edge.
Integrate-and-fire neurons, shown in Figure~\ref{fig:neuron}, are
adopted in this work.  Each of these neurons is composed of three
elements: (i) an integrator, (ii) a memory (containing its current
\emph{activation state}); and (iii) a non-linear transfer function (a
hard-limiter is adopted henceforth).  A spike is produced whenever
the activation state equals or exceeds the respective threshold $T(i)$.
In this work, all states and spikes are updated synchronously at every
time step $t$.  Once a fire takes place, the internally accumulated
activation is distributed equally amongst the outgoing axons.
Though such a distribution is not biologically realistic, it can be
emulated by considering varying synaptic weights.  The distribution of
the activation implies in the conservation~\cite{Costa_nrn:2008,
Costa_begin:2008, Costa_activ:2008} of the total stimuli conveyed by
the source node.

Though all the 7 models of theoretical networks adopted in this work
are, in principle, undirected, they can immediately used to obtain
directed complex neuronal networks by splitting each undirected edge
into two respective directed edges.  Now, for each cell the incoming
edges represent the synapses and the outgoing edges the axons.
Consequently, the in- and out-degrees are equal for every cell.  The
activation of such networks is henceforth obtained by placing a source
of constant activation (intensity 1) at a specific nodes and
monitoring the respective activation and spiking patterns along
time.

\section{Extreme Configurations}

\begin{figure}[htb] 
  \begin{center}
  \includegraphics[width=1\linewidth]{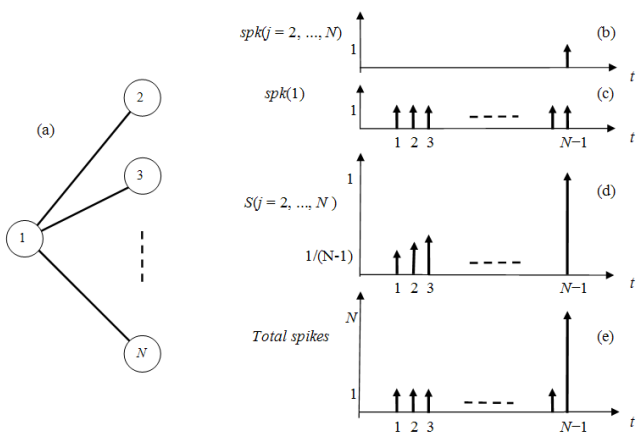} 
  \caption{A simple network (a) containing a hub (node 1) and
              the respective diagrams of spikes at any of the
              neurons 2 to $N$ along $t$ (b), spikes at the source
              neuron 1 (c), activations at any of the neurons
              2 to $N$ (d), and the total number of spikes in the
              network (e).
  }~\label{fig:hub} 
  \end{center}
\end{figure}

\begin{figure*}[htb] 
  \begin{center}
  \includegraphics[width=0.5\linewidth]{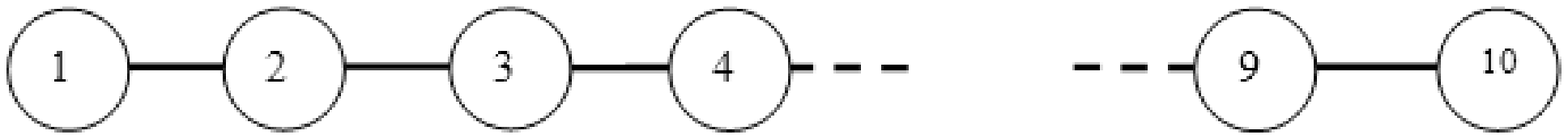} (a)  \\  \vspace{0.5cm}
  \includegraphics[width=0.4\linewidth]{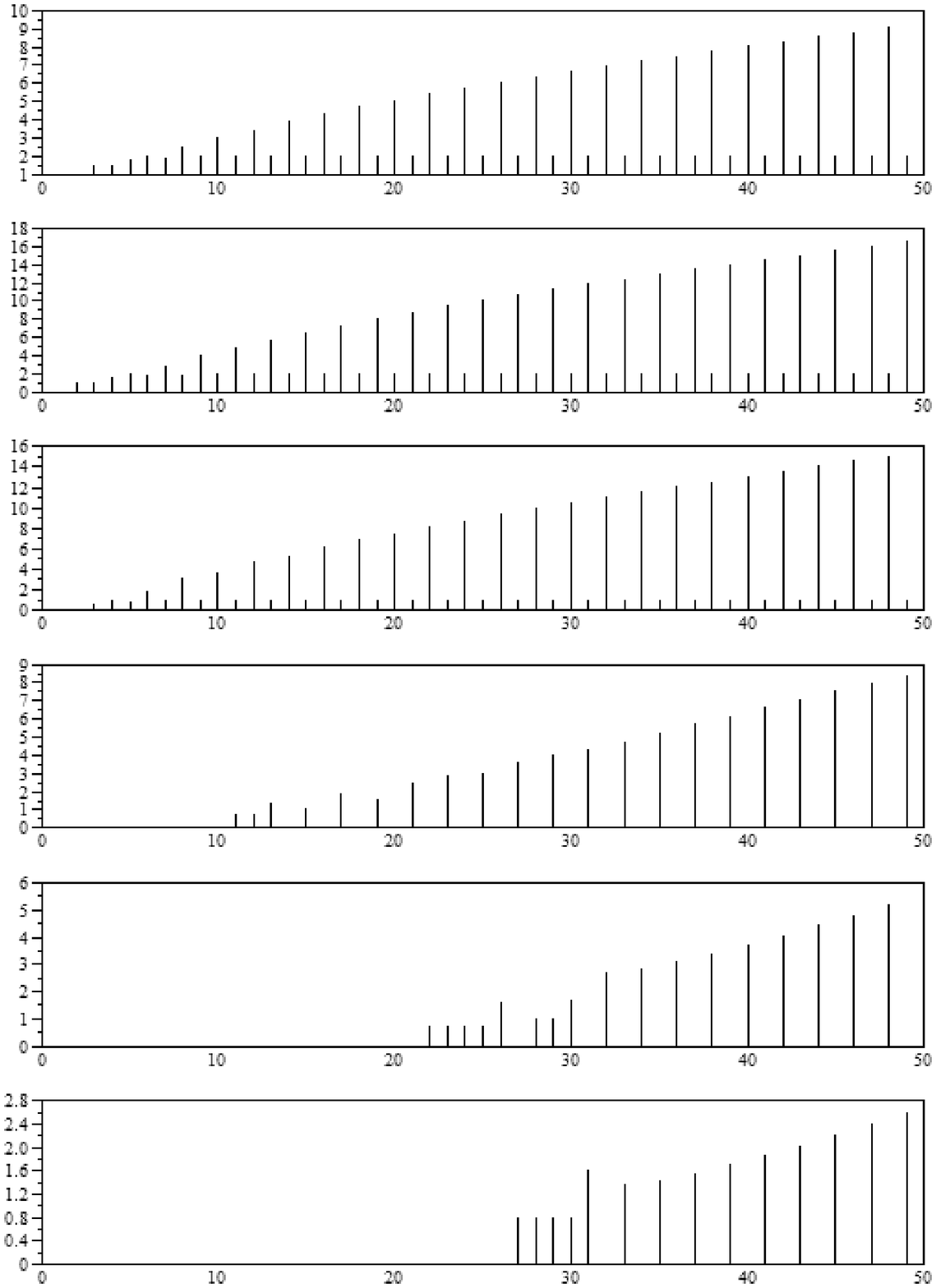} 
  \includegraphics[width=0.45\linewidth]{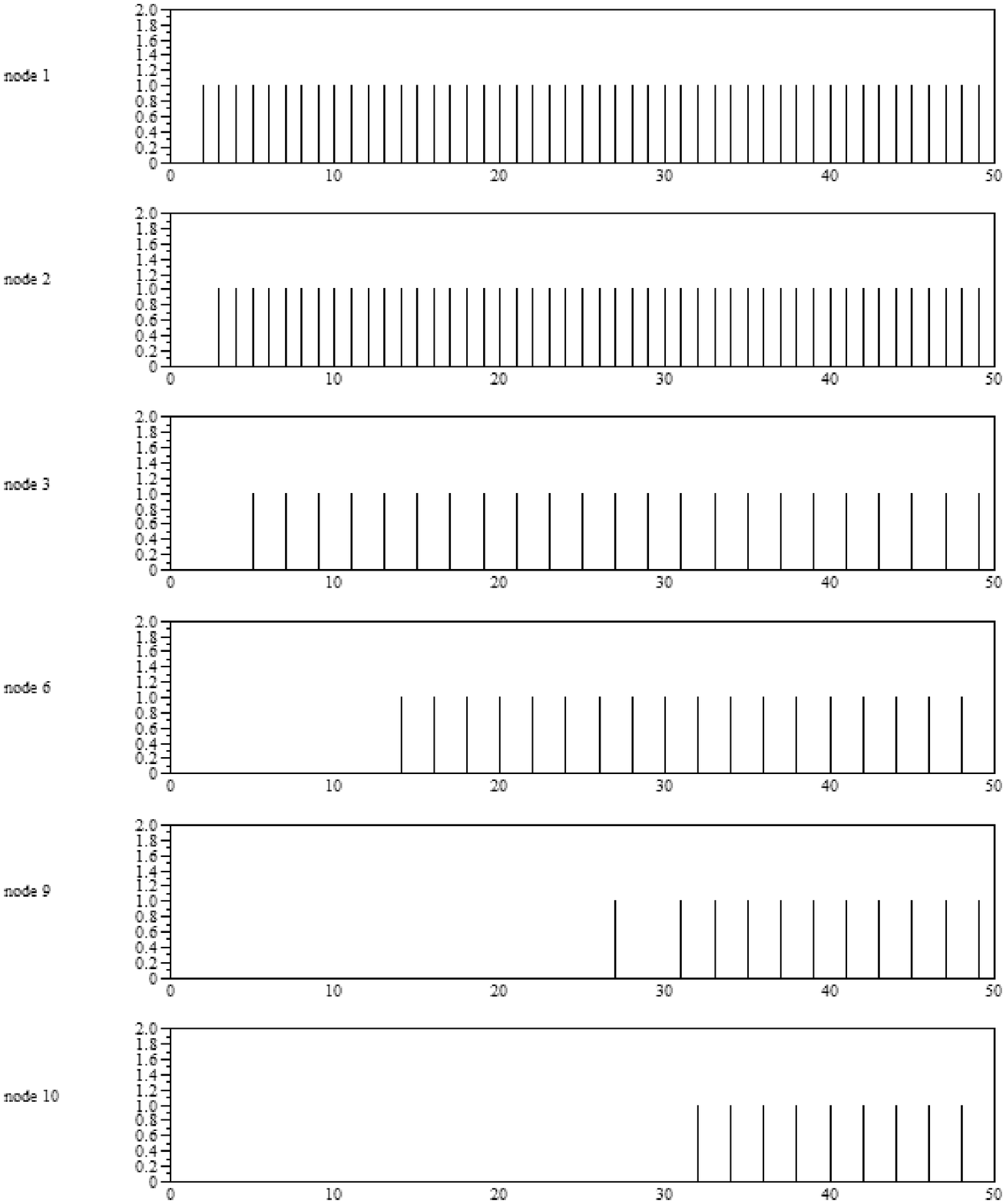}  \\
  (b)   \hspace{8cm} (c)
  \caption{The chain network (a) and the activations (b) and spikes (c)
                    obtained for several of its nodes (identified in
                    the middle of the picture).  All activation
                    patterns in (b) exhibit the saw oscilations, which
                    also alternate between subsequent nodes (a phase
                    shift by one time step).  After a brief
                    transient, the peaks in these saw patterns tend to
                    increase almost linearly with time, with different
                    slopes.  Quite regular spiking
                    patterns can be seen in (c).
  }~\label{fig:chain} 
  \end{center}
\end{figure*}

In this section, we investigate the activation spread in two extreme
types of complex neuronal networks, namely the hub and a chain dual
structures~\cite{Costa_path:2007} shown in Figures~\ref{fig:hub}
and~\ref{fig:chain}.  Such examples are aimed at familiarizing the
reader with the adopted concepts and representation while illustrating
how the activation disseminates in two extreme, but particularly
important, situations.

Let us start by analysing the situation depicted in
Figure~\ref{fig:hub}.  Here, the network incorporates 10 nodes, one of
each (node 1) is adopted as the source.  Observe that node 1 is the
hub of this simple network.  All neurons have the same threshold
$T=1$.  After activation of the source (kept constant and equal to 1)
at $t= 1$, neuron 1 starts firing at every time step $t$, as shown in
Fig~\ref{fig:hub}(c).  Such spikings activate the neurons attached to
neuron $i$ by an amount of $1/(N-1)$ at every step, implying linear
accumulation of the activation along time inside each of the neurons 2
to $N$ (Fig.~\ref{fig:hub}(d).  At time $t=N-1$, such accumulated
activations will reach the threshold $T=1$, implying the simultaneous
spiking of all neurons from 2 to $N$.  Because neuron 1 spikes
continuously, we have a total of $N$ spikes at time $t=N-1$.  This
simple example illustrates the basic effect underlying the avalanches
of activation in the complex neuronal networks.  However, such
avalanches are \emph{by no means exclusively related to the firing of
hubs}.  First, most of the networks exhibiting activation avalanches,
including the BA model, incorporate nodes with varying degrees.
Second, avalanches are found in completely degree-regular networks
such as the PI model, being also pronounced for highly degree-regular
networks such as ER and PN.  As shown in this work, the avalanches are
ultimately a consequence of hubs of \emph{hierarchical degree} along
the concentric organization of the network with respect to the
source node.

\begin{figure}[htb] 
  \begin{center}
  \includegraphics[width=0.8\linewidth]{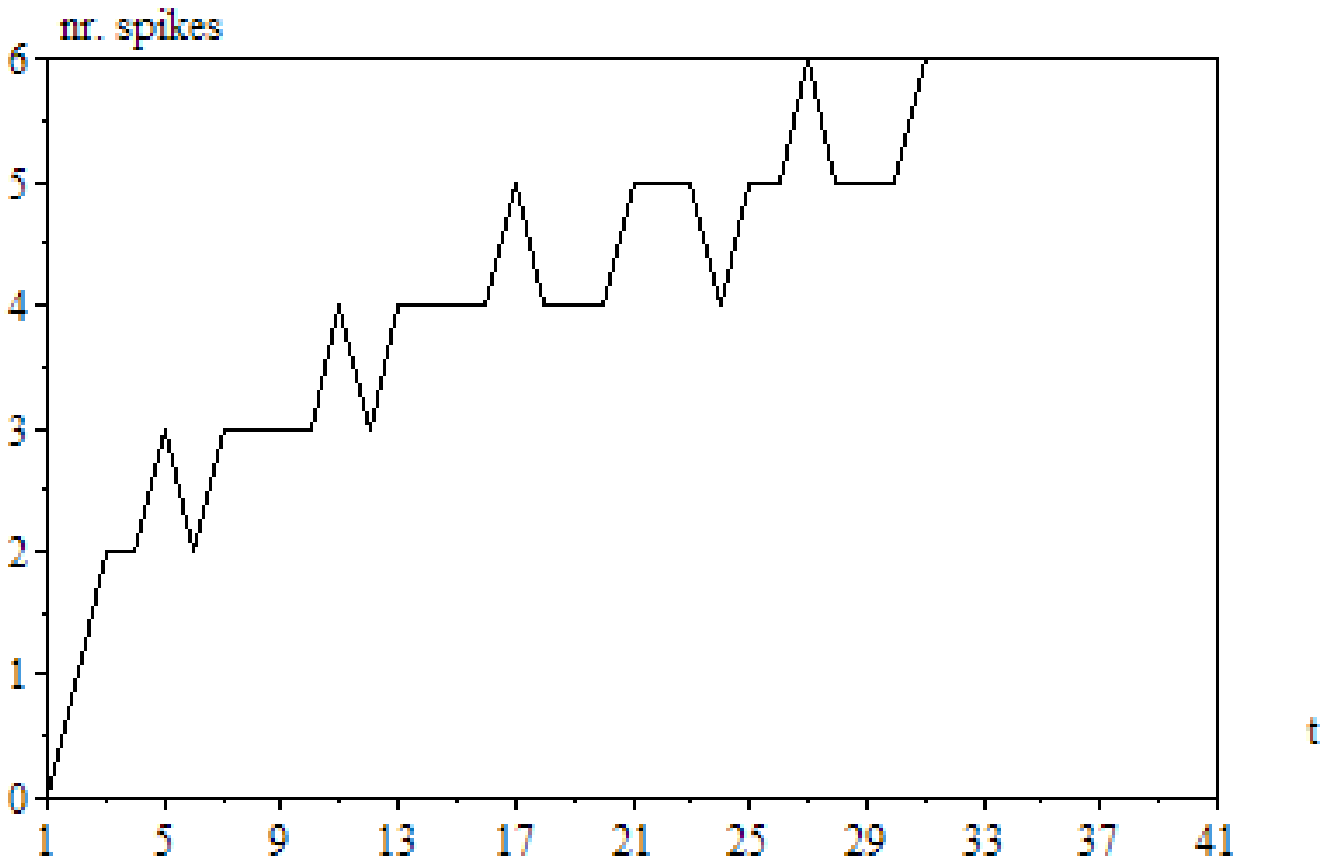} 
  \caption{The total of number of spikes generated along time by
                the chain network in Figure~\ref{fig:chain}.
                The activation is transferred gradual and slowly
                from the nodes in the left-hand side to those in
                the right-hand side of the chain network.
  }~\label{fig:chain_spks} 
  \end{center}
\end{figure}

The network illustrated in Figure~\ref{fig:chain}(a) is the
dual~\cite{Costa_path:2007} of a hub, i.e. it is completely organized
as a \emph{chain}~\cite{Boas:2007} of nodes.  Despite the simplicity
of this network, the activation dynamics is relatively involved
because of the backward activation which takes place at each firing of
cells 2 to N.  The activations and spikings along time for neurons 1,
2, 3, 6, 9 and 10 are shown in Figure~\ref{fig:chain}(b) and (c),
respectively.  All activations in this figure exhibit what is
henceforth called \emph{saw oscillations}, i.e. alternate high and low
values, separated by one time step.  The high values of the saw
oscillations are henceforth called \emph{peaks}.  The peaks obtained
for each node tend to undergo, after a brief transient period, an
almost linear increase with time with varying slopes.  By comparing
the several diagrams in Figure~\ref{fig:chain}, it becomes clear that
the activation is propagated gradual and progressively to the neurons
at the right-hand side of the chain network, with the initial neurons
concentrating activation because of the backward flow of activation.
Figure~\ref{fig:chain_spks} shows the total number of spikes generated
along time by the chain network in Figure~\ref{fig:chain}.  The smooth
and gradual distribution of the activation of the chain networks can
be clearly appreciated from these results.

The saw oscillation can be better understood by considering the
following coupled difference equations involving 5 consecutive nodes
$i-2$ to $i+2$ centered at a node $i$ along the chain (except at the
two extremities)

\begin{eqnarray}
 S(i-1) = 0.5 S(i-2) \delta_{i-2} + S(i-1) + 0.5 S(i) \delta_{i} \nonumber \\
 S(i) = 0.5 S(i-1) \delta_{i-1} + S(i)   + 0.5 S(i+1) \delta_{i+1} \nonumber \\
 S(i+1) = 0.5 S(i)   \delta_{i}   + S(i+1) + 0.5 S(i+2) \delta_{i+2} \nonumber \\ \nonumber
\end{eqnarray}

where $S(i)$ is the accumulated state of node $i$ and $\delta_{i}$ is
the function specifying the time when neuron $i$ fires.  Observe that
these functions are not easily obtained for most cases.  Whenever one
of such functions $\delta_{v}$ is activated, corresponding to the
respective spiking, the internal state of the respective neuron $v$ is
flushed out, being distributed equally among its left and right
neighbors, and subsequently cleared.

Let us assume that the saw oscillation is verified for the current
instant, i.e. we have a high value (above threshold) at neuron $i-2$,
a low value (below threshold) at neuron $i-1$, and to on
alternately. Because neurons $i-2$, $i$ and $i+2$ have internal states
larger than the threshold $T=1$, they spike, and their activations are
shared between their neighbors.  Because the latter were at a low
activation, they do not send any substantial input to any of the nodes
$i-2$, $i$ and $i+2$, which therefore become lowly activated.
However, the high activations previously contained in these three
neurons are now transferred to neurons $i-1$ and $i+1$, which become
highly activated.  An opposite exchange is verified for the subsequent
step, and so on, corroborating the stability of the attractor implying
the chain oscillations.

In spite of the relatively complex shapes of the activations along
time, quite regular spiking patterns are obtained, as shown in
Figure~\ref{fig:chain_spks}(c).

\section{Concentric Characterization of the Topology of the Network Models}

This section presents the concentric characterization of the 7 types
of complex networks considered in this work.  First, we estimate
experimentally the distribution of the hierarchical number of nodes
$n_h(i)$, hierarchical degree $k_h(i)$, and intra-ring degrees
$a_h(i)$ by averaging among every node $i$.  Each of the 7 theoretical
models of complex networks considered in this work had their
hierarchical measurements (hierarchical number of nodes, hierarchical
degrees, and intra-ring degrees) obtained computationally for each
node of representative samples of each of the topological types.
Because systematic comparisons of hierarchical measurements of diverse
complex networks models have been reported previously
(e.g.~\cite{Costa:2004, Costa_NJP:2007, Costa_JSP:2006,
Costa_EPJB:2005}), this section is aimed only at illustrating the
overall hierarchical structures of the considered network models.

\begin{figure*}[htb] 
  \begin{center}
  \includegraphics[width=0.8\linewidth]{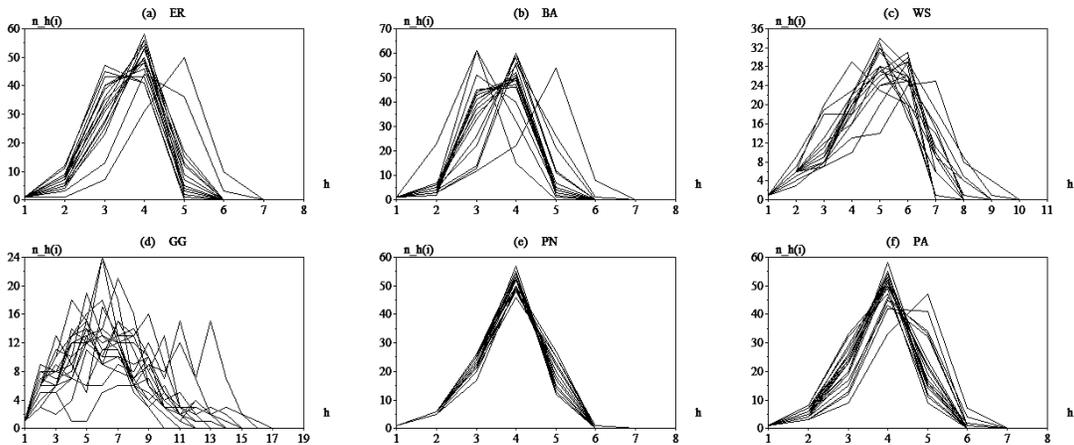} 
  \caption{The hierarchical number of nodes obtained considering 20  
                  nodes chosen at random from each of the 6 considered
                  networks.
  }~\label{fig:nnodes} 
  \end{center}
\end{figure*}

Figure~\ref{fig:nnodes} shows the hierarchical number of nodes
$n_h(i)$ obtained for 20 nodes randomly chosen from each of the
networks of different types.  In all cases, the hierarchical node
degree starts with value one (the initial node, at $c=0$) and
increases until reaching one or more (in the case of the GG structure)
peaks, decreasing subsequently to zero (the end of the network).
Interestingly, the peaks for most models tend to occur near the middle
hierarchical level.

However, the specific details of the obtained signatures are directly
related to the topology of each type of network.  The most regular
distribution of hierarchical number of nodes was obtained for the PN
model which, after the PI structure, is the most structurally regular
model among the types of networks considered presently.  Homogeneous
signatures were obtained also for the ER and PA models.  The GG
structure (Fig.~\ref{fig:nnodes}d) implies not only the largest
diversity of signatures, but also the longest signatures (compare the
$x-$axes ranges in each plot in Figure~\ref{fig:nnodes}).  The
second longest signatures were obtained for the WS network
(Fig.~\ref{fig:nnodes}c).  Because the sum of the hierarchical number
of nodes needs to be constant and equal to the total number of nodes
$N$, the two cases leading to the longest signatures also imply the
lowest values in the $y-$axes.  

\begin{figure*}[htb] 
  \begin{center}
  \includegraphics[width=0.8\linewidth]{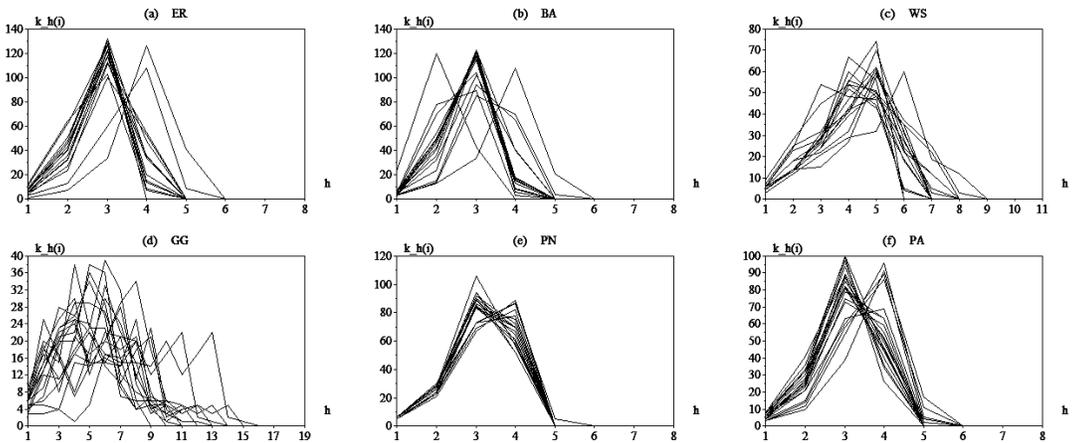} 
  \caption{The hierarchical degrees obtained considering 20  
                  nodes chosen at random from each of the 6 considered
                  networks.
  }~\label{fig:hdeg} 
  \end{center}
\end{figure*}

The hierarchical degrees $k_h(i)$ obtained for each of the types of
networks are shown in Figure~\ref{fig:hdeg}.  Recall that the
hierarchical degree of a node $i$ at concentric level $h$ corresponds
to the number of edges between the $h-$ and $(h+1)-$neighbors of
$i$. Again, the hierarchical degrees tend to increase to a peak,
decreasing thereafter, with the exception of the GG structure
(Fig.~\ref{fig:hdeg}d).  The PN yielded the most similar signatures,
reflecting its enhanced structural regularity.  The longest and lowest
signatures were again obtained for the GG and WS models.  Observe that
the peak of the hierarchical degrees tend to occur one level before
that where the maximum number of nodes are obtained.  

\begin{figure*}[htb] 
  \begin{center}
  \includegraphics[width=0.8\linewidth]{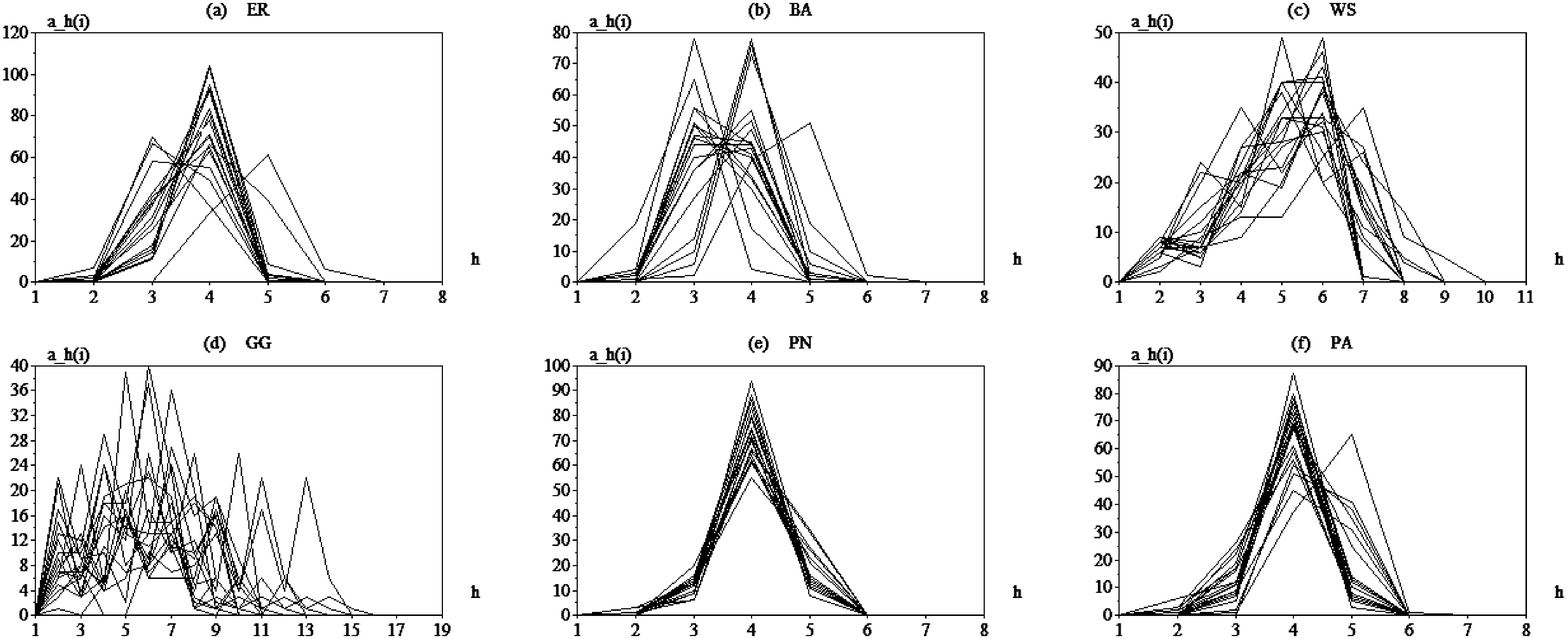} 
  \caption{The hierarchical intra-ring degrees obtained considering 20  
                  nodes chosen at random from each of the 6 considered
                  networks.
  }~\label{fig:intra} 
  \end{center}
\end{figure*}

Figure~\ref{fig:intra} shows the hierarchical intra-ring degrees
$a_h(i)$ obtained for each network type.  These signatures exhibit
properties similar to those discussed above.

\begin{figure}[htb] 
  \begin{center}
  \includegraphics[width=0.6\linewidth]{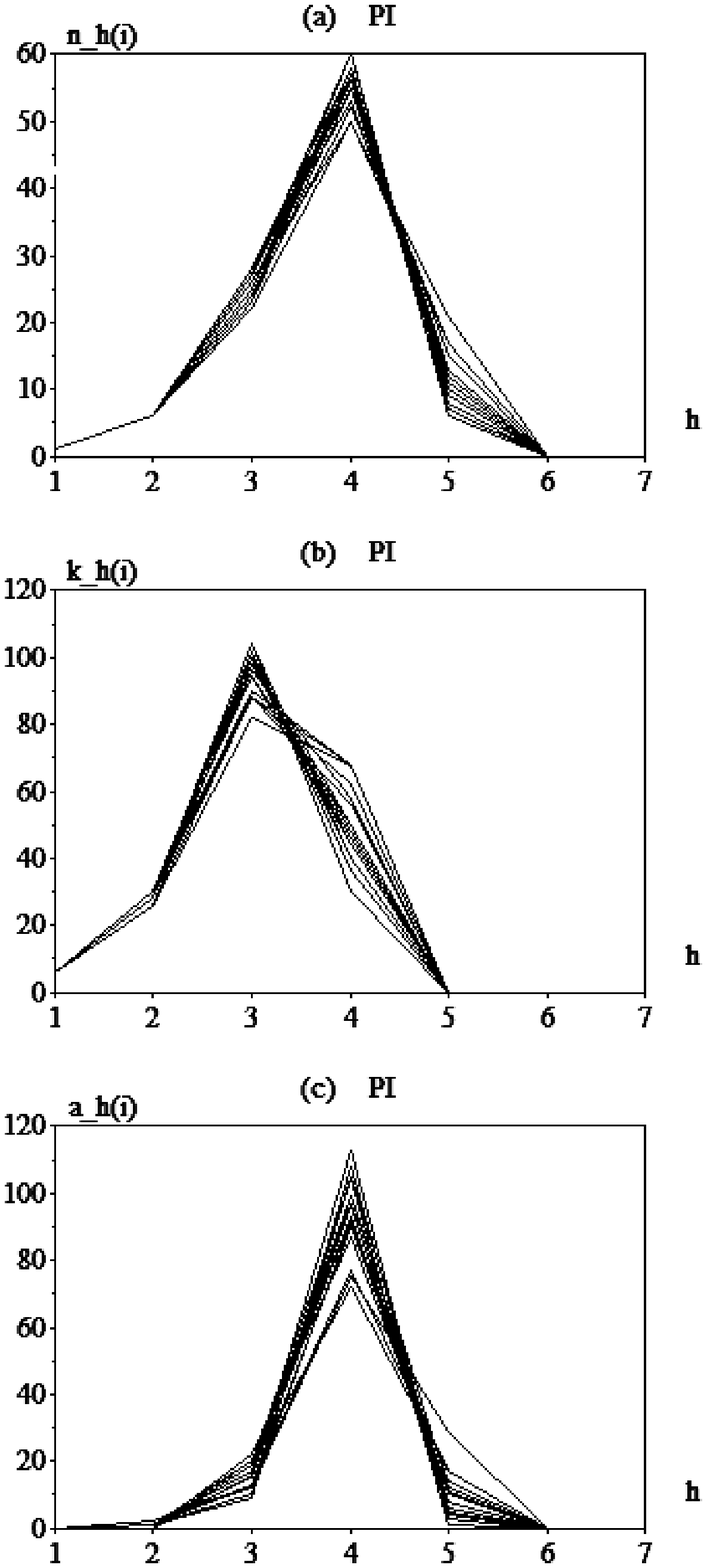} 
  \caption{The hierarchical number of nodes (a), hierarchical degrees (b),
                 and intra-ring degrees (c) obtained for the PI
                 network model.
  }~\label{fig:conc_PI} 
  \end{center}
\end{figure}

Figure~\ref{fig:conc_PI} shows the hierarchical number of nodes (a),
hierarchical degrees (b) and intra-ring degrees (c) obtained for the
highly regular PI network.  As expected, they yielded the most uniform
sets of hierarchical signatures.

\section{The Equivalent Model:  Connecting the Hierarchical 
Structure and Transient Dynamics} \label{sec:conc}

\begin{figure*}[htb] 
  \begin{center}
  \includegraphics[width=0.7\linewidth]{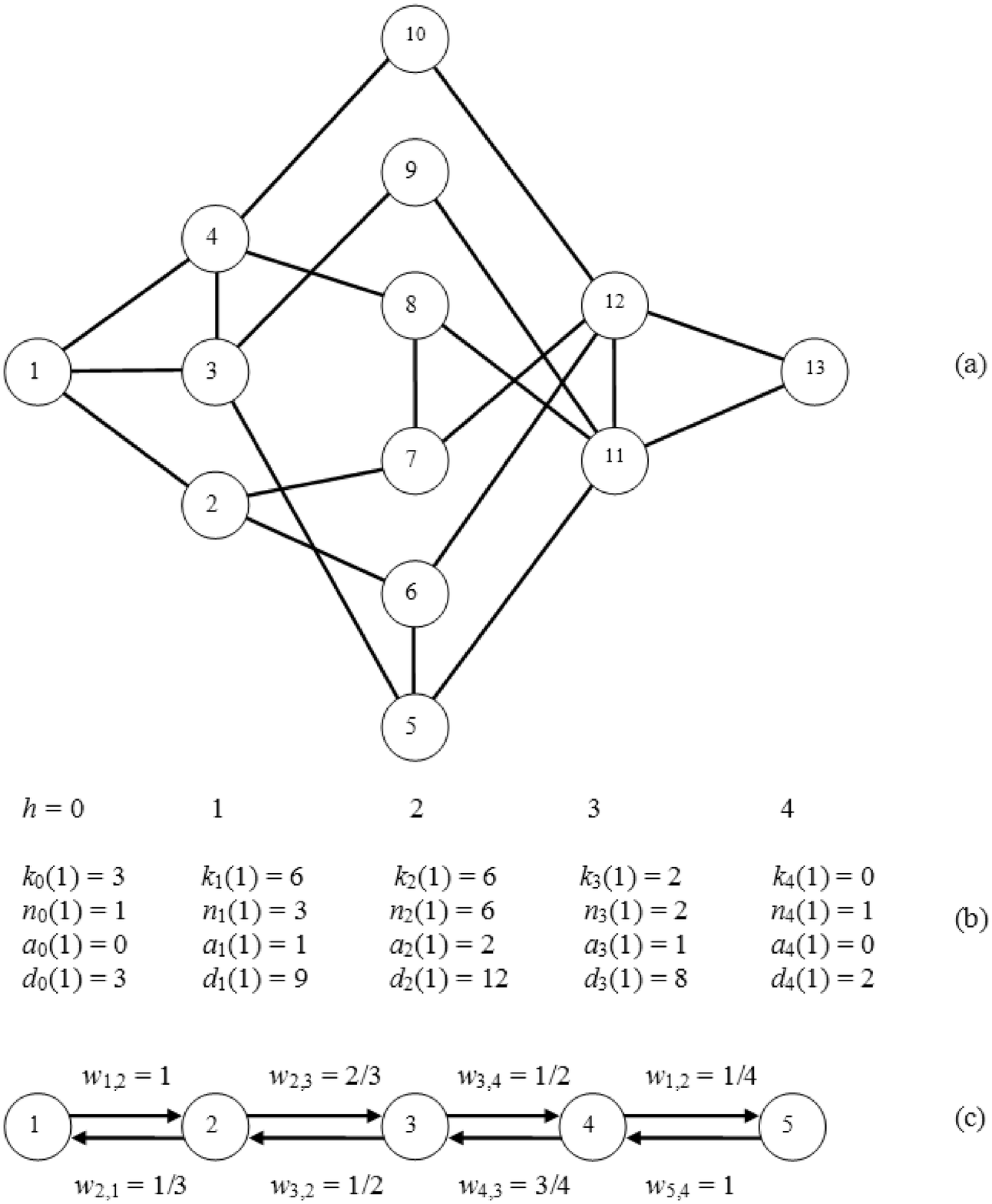} 
  \caption{An example of equivalent model of a network (a).  Node
              1 is chosen as the reference, defining
              5 concentric levels $c = 0, 1, \ldots, 4$.  The 
              respective hierarchical measurements are shown in (b).
  }~\label{fig:equiv} 
  \end{center}
\end{figure*}

The hierarchical (or concentric) organization of complex networks with
respect to a reference node provides not only additional information
about the network connectivity at successive larger topological scales
around the reference node (a multiscale approach), but is also
intrinsically related to the dynamics of distribution through the
network of activation received from the reference node.  In order to
discuss this important fact in more detail, let us consider the simple
network in Figure~\ref{fig:equiv}, containing 13 nodes.  Node 1 was
taken as the reference node, and the 5 hierarchical neighborhoods
identified, being shown as columns in Figure~\ref{fig:equiv}.  The
hierarchical number of nodes, hierarchical degrees and intra-ring
degrees of node 1 are shown below the network.  Also shown are the
\emph{total inter-ring degrees} of each $h-$neighborhood, obtained by
adding its respective hierarchical degree and the hierarchical degree
at the previous level, i.e. $d_h(i) = k_h(i) + k_{h-1}(i)$.  It is
interesting to observe that, though small, this network captures the
hierarchical organization typically found in complex networks, namely
the peak of the hierarchical measurements, occurring near the
intermediate concentric level.

Let us now suppose that self-avoiding random walks start continuously
at the reference node 1.  The total number of different paths which
can be followed by the walks (a measurement related to the diversity
of random walks~\cite{Costa_diverse:2008, Latora_entropy,
Gardenes:2007}) can be estimated by multiplying the hierarchical
degrees at each level, except for the very last one.  The estimation
would be exact were not for the intra-ring edges, which allow the
moving agents to return to previous levels.  In addition, the average
time it takes to activate nodes in a given concentric level $c$ is
readily given as being very close to $h$.  Therefore, the concentric
organization of a network with respect to the reference node can yield
important information not only about the total number of different
walks, but also about the time at which nodes receive their first
non-zero activation.  In the specific case of self-avoiding random
walks (or even traditional random walks), the activation of the nodes
proceeds gradually, without any abrupt increases or decreases along
time.

Let us now consider how the network in Figure~\ref{fig:equiv} behaves
with respect to the integrate-and-fire non-linear dynamics adopted in
the current work.  The source of activation is placed at node 1, and
the activation is propagated every time any of the neurons fires.
Because the activation being sent from node 1 to nodes 2, 3 and 4 is
equally distributed among the 3 respective axons, i.e. edges $(1,2)$,
$(1,3)$ and $(1,4)$, it takes three steps for the neurons in level 1
to fire.  However, when they fire they do so simultaneously,
generating a small surge of number of spikes.  As each of these three
neurons fires, they sends equal shares of activation to all neurons to
which they are connected.  For instance, after firing, neuron 2
distributes its activation among neurons 1, 6 and 7.  Because all
neurons in the concentric level 1 send connections to more neurons in
the next level, those neurons receive only a small share of the
activations sent by the neurons in level 1.

An opposite effect is observed when the neurons at level 2 fire.  Now,
because the neurons at level 3 receive three axons each from neurons
in the previous level, a substantial amount of activation is received
by each of them.  With few exceptions related to intra-ring edges and
degree non-homogeneities, the neurons in the next level will
consequently fire at the next time step because of the intense
activation received from the previous level.  Therefore, once the
neurons in the level with highest hierarchical number of neurons
spike, they tend to imply intense activation to both the previous and
following levels.  Now, considering that the neurons at the previous
levels are already well-activated as a consequence of being closer to
the source, and because the neurons in the following layers will
receive intense activation from the spiking neurons,
\emph{the spiking of the neurons in the concentric level with the
highest hierarchical number of nodes tends to imply the spiking of
most (often all) the other neurons in the network} within a one or a
few time steps.  This corresponds to how the avalanches of spikings
are produced in typical complex neuronal networks.

It should be observed that the regularity of connections of all
neurons in Figure~\ref{fig:equiv} was deliberately chosen for the sake
of a more didactic presentation.  Non-homogeneities of connections
between nodes (e.g. a neuron in level $h$ send more connections to
distinct neurons in the next level than other neurons at level $h$)
imply in non-simultaneous spiking of the neurons at a given level.
However, because several of the 7 considered network types are
characterized by strong degree regularity (the main exception is the
BA case), they degree homogeneity is not a particularly restrictive
assumption.  Another simplification adopted so far concerns the effect
of intra-ring connections in slightly delaying the activation of the
adjacent levels to a level $h$, by implying a fraction of the
activation liberated by the spiking neurons at level $h$ to remain at
that same level.  Intra-ring connections can also imply in small lost
of simultaneity in the spiking of the neurons at the next level.
Though all the developments reported in the remainder of this article
can be complement to take into account non-homogeneous degree
distributions and intra-ring connections, such extensions are left for
future works.

Because of the relative uniformity of the degree distributions in most
of the 7 considered complex network models, it is possible to reduce
the hierarchical organization of a complex, with reference to a given
node, into a chain network (see Fig.~\ref{fig:equiv}c), yielding a
mean-field \emph{equivalent model} of the relationship between the
hierarchical structure and the non-linear activation dynamics in
complex networks.  The basic idea is to consider the activation and
spiking of all neurons in each concentric level to be subsumed by a
single equivalent node.  Therefore, an equivalent chain network such
as that in Figure~\ref{fig:equiv} is obtained from the original
network.  Now, the connections between these equivalent nodes need to
be \emph{weighted} in order to account to the typically asymmetric
distribution of activation between the nodes at the previous and next
concentric levels.  These weights are properly defined as

\begin{eqnarray}
  w_{h,h+1} = \frac{k_h(i)}{d_h(i)}  \nonumber \\
  w_{h-1,h} = \frac{k_{h-1}(i)}{d_h(i)}  \nonumber
\end{eqnarray}

where $c = 2, 3, \ldots, h_{max}-1$.  In the particular case of the
neurons at the extremities of the chain, we have

\begin{eqnarray}
  w_{0,1} = 1  \nonumber \\
  w_{h_{max},h_{max}-1} = 1  \nonumber
\end{eqnarray}

\begin{figure*}[htb] 
  \begin{center}
  \includegraphics[width=0.7\linewidth]{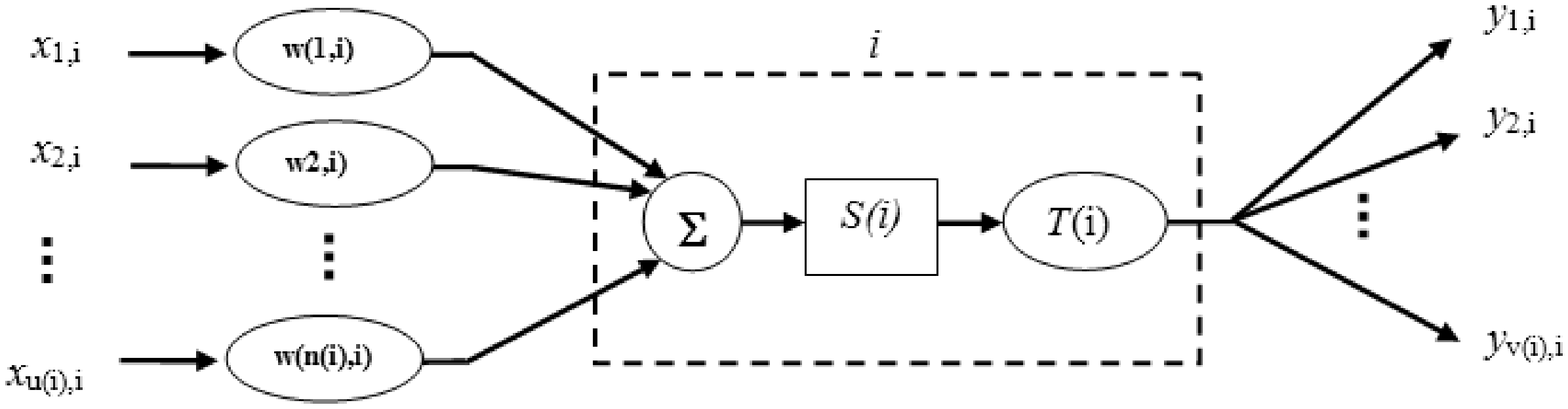} 
  \caption{The integrate-and-fire neuron with weights.
  }~\label{fig:neuron_w} 
  \end{center}
\end{figure*}

Figure~\ref{fig:neuron_w} shows the adopted integrate-and-fire
neuronal model modified to incorporate weights.

In order to reflect the division of the activation received at a level
among all its constituent neurons, it is necessary to set the
thresholds of the neurons in the chain equivalent neuronal network as
being equal to the respective hierarchical number of nodes, i.e. $T(h)
= n_h(i)$.  This last modification completes the definition of the
mean-field equivalent model of the non-linear distribution of
activation through reasonably degree-regular complex neuronal
networks.  

Once the equivalent model of a given complex network and activation
source has been obtained from the respective concentric properties,
the following important estimations about the activation/spiking
dynamics can be obtained:

{bf Avalanche hierarchy: } The concentric level $c$ at which the
avalanche initiates can be estimated as the level which has the
largest number of neurons.

{\bf Beginning of Activation Times:} The time it takes to a level to
spike can be estimated by adding the number of nodes from level 0 to
the current level.

{\bf Avalanche times: } The time at which the first avalanche will
occur can be estimated by adding the number of nodes from level $1$ to the level containing the largest number of nodes (inclusive).  

{\bf Intensity of avalanches: } The intensity of the avalanche at the
time it occurs can be estimated as the number of nodes in the critical
level.

{\bf The evolution of the avalanches: } Though the maximum variation
of the total number of spikes is approximately given as above, varying
degrees and intra-ring connections will imply in a dispersion of the
spiking, affecting the intensity and width of the avalanches.  Such
additional features can be estimated by taking into account the number
of nodes at concentric levels which are adjacent to the critical level
(i.e. that containing the largest number of nodes).

In the following, the potential of the equivalent model for predicting
the properties of the avalanches is illustrated with respect to the
simple network in Figure~\ref{fig:equiv} and then with respect to each
of the 7 topological types of networks adopted in this article.

\section{Illustrative Examples}

\begin{figure*}[htb] 
  \begin{center}
  \includegraphics[width=0.39\linewidth]{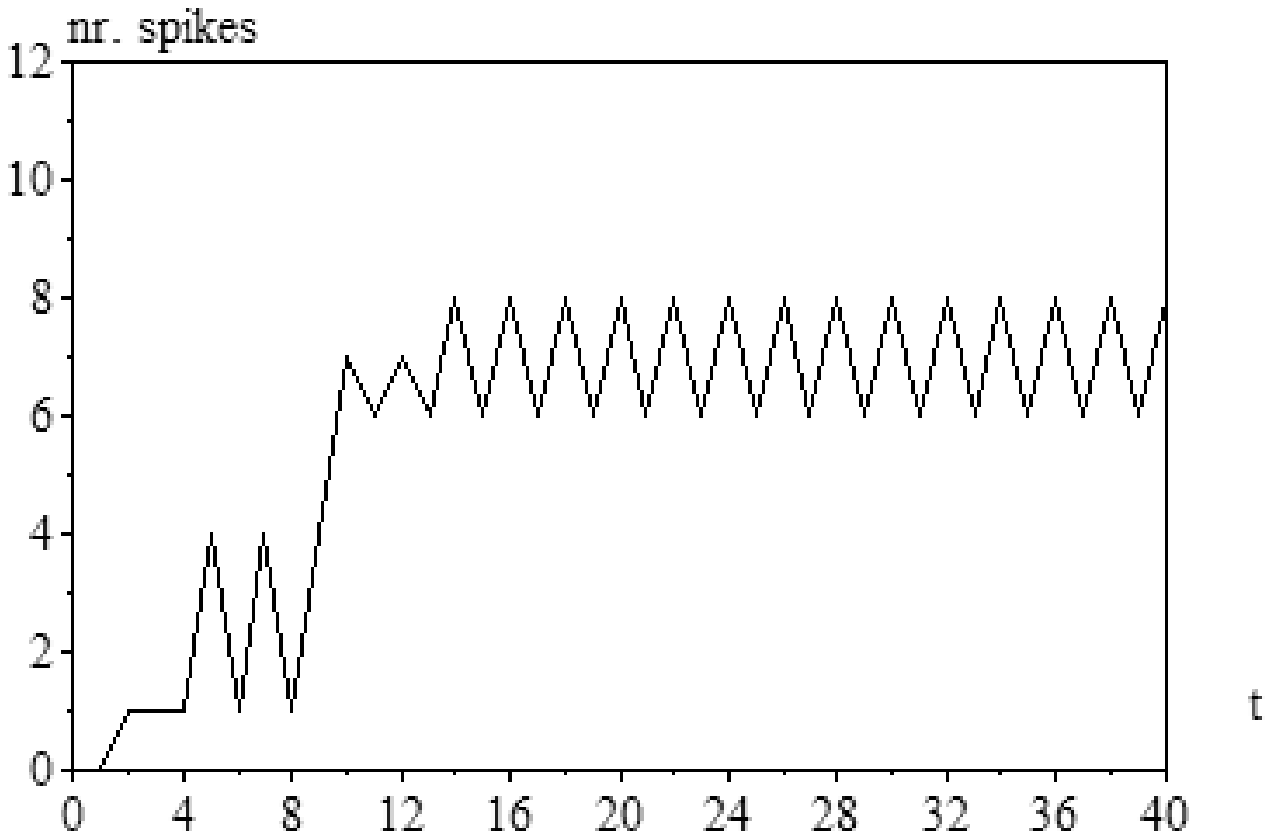} 
  \includegraphics[width=0.39\linewidth]{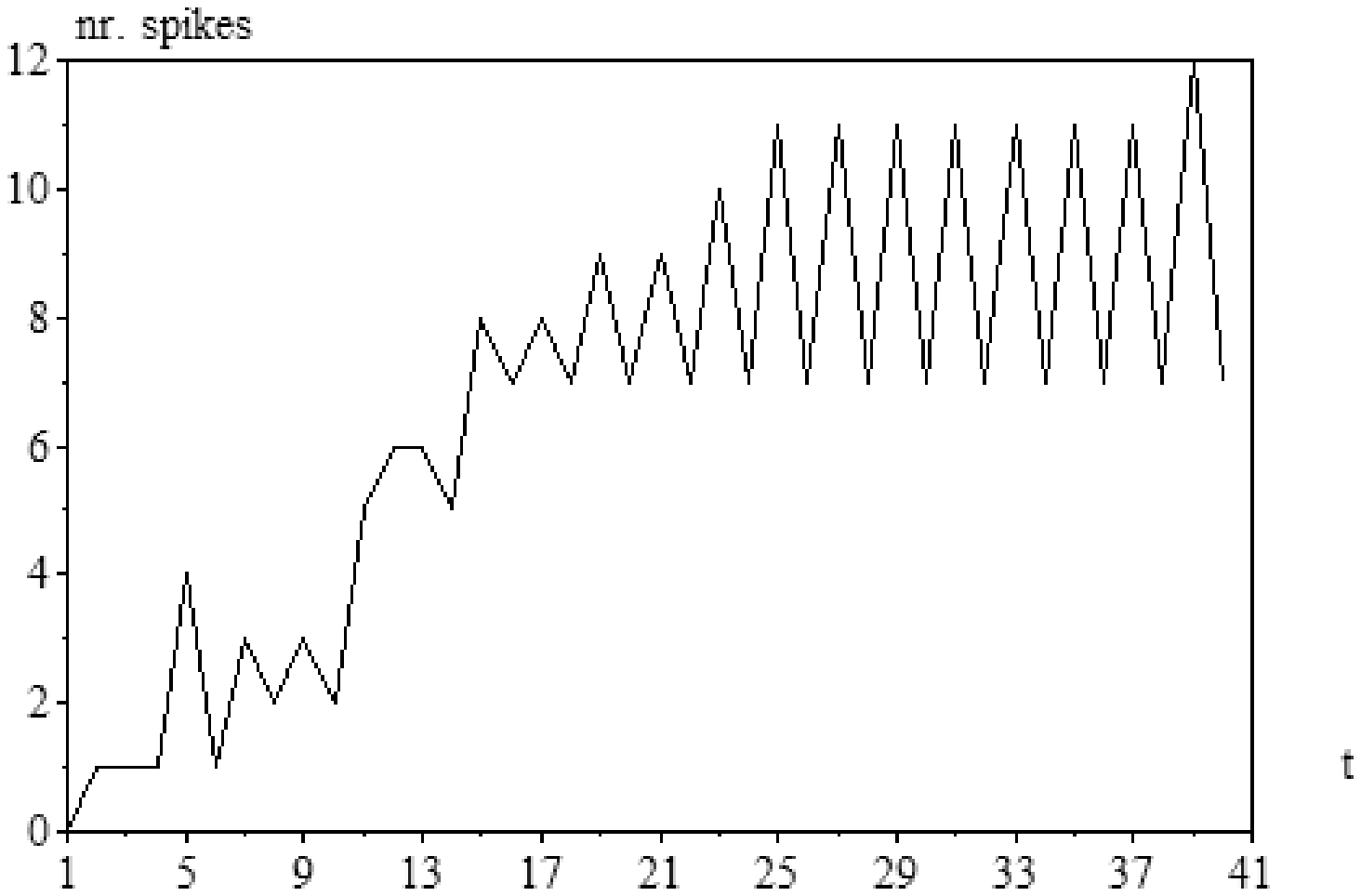} \\
  (a) \hspace{7cm} (b)
  \caption{The total number of spikes along time obtained for the
                network in Figure~\ref{fig:equiv}(c) by considering
                the whole original structure (a) and the respective
                equivalent model (b).  The small differences between
                these two curves are a consequence of the intra-ring
                degrees, which are not considered in the equivalent 
                model.  
  }~\label{fig:ex_simpl} 
  \end{center}
\end{figure*}

Figure~\ref{fig:ex_simpl} shows the total number of spikes along time
obtained for the complete original network (a) as well as by
considering the respective equivalent model in
Figure~ref{fig:equiv}(c).  Identical curves are obtained in case the
intra-ring connections are removed from the network.  However, the
fact that these connections are not considered in the equivalent model
implies in the predicted slight delays and spread of the activations.
Nevertheless, the overall dynamics leading to the spiking avalanche
was properly captured and reproduced by the model up to the small
differences implied by the elimination of the intra-ring
connections.

\begin{figure*}[htb] 
  \begin{center}
  \includegraphics[width=0.9\linewidth]{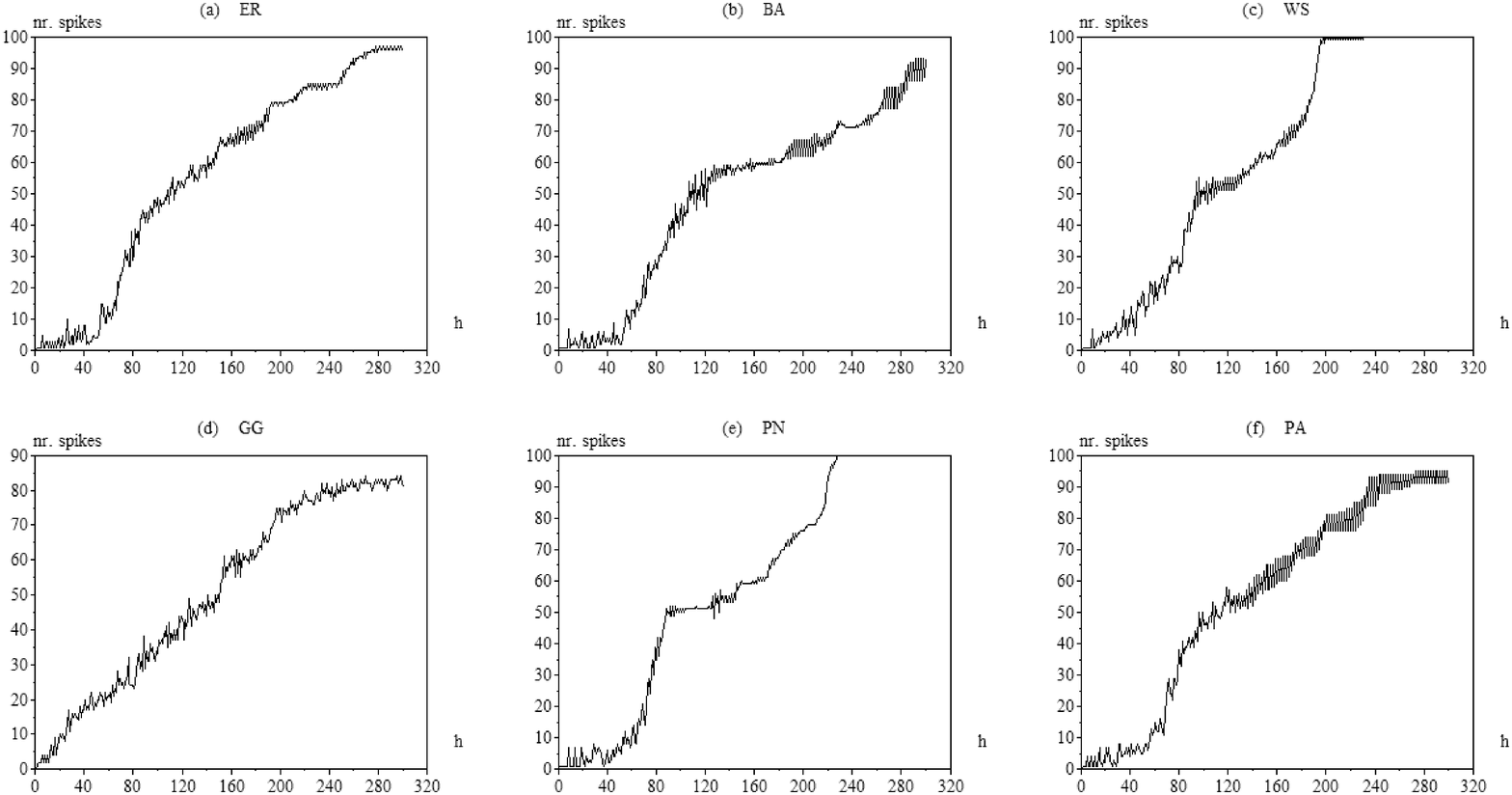} 
  \caption{The total number of spikes along time obtained for the
             ER, BA, WS, PN and PA models considering the whole
             original networks.
  }~\label{fig:res_orig} 
  \end{center}
\end{figure*}

Figure~\ref{fig:res_orig} shows the total number of spikes in terms of the
time steps, as obtained by considering the whole ER, BA, WS, GG, PN
and PA complex network models adopted in the present work with the
activation source placed at node 50.  The most remarkable feature of
such results, which soon catches our attention while looking at
Figure~\ref{fig:res_orig}, is that most networks, except the GG but
including the BA, yielded very similar patterns of total spiking along
time.  Moreover, all networks (except GG) show avalanches, which occur
at similar times near 80 steps.  Similar results have been obtained
for most of the other nodes, but are not shown here for the sake of
space.  These results suggest that avalanches, as well as their onset
time, \emph{seem to be an almost universal feature of complex neuronal
networks}.  The only parameters which are likely to affect the
avalanche parameters are the sizes and average degrees of the
networks.  Because of its potential importance, we investigate this
issue further in the next section.  Also of particular interest in
Figure~\ref{fig:res_orig} are the occurrence of a second avalanche in
the WS and PN cases.

\begin{figure*}[htb] 
  \begin{center}
  \includegraphics[width=0.9\linewidth]{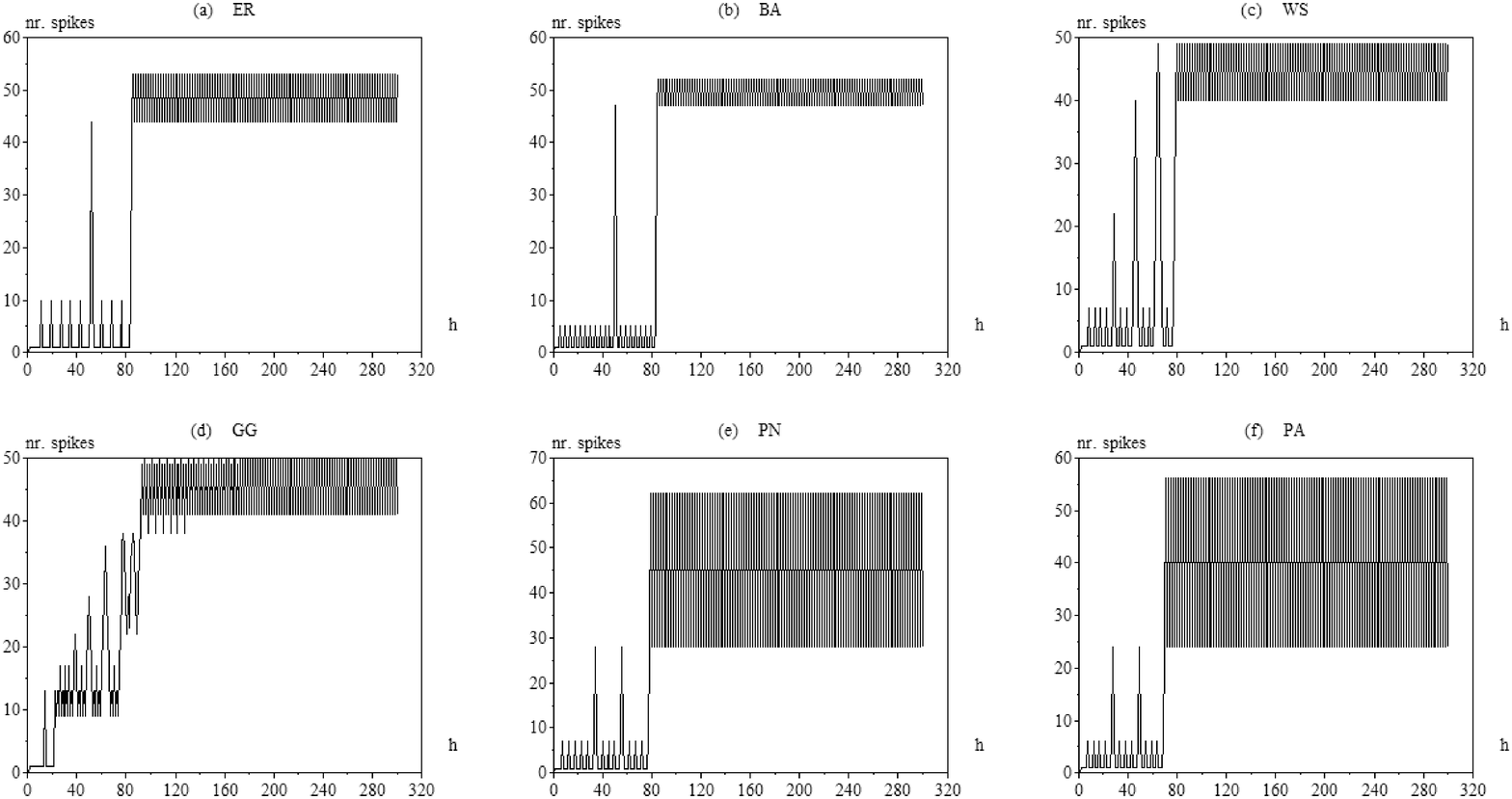} 
  \caption{The total of spikes in terms of time obtained by 
                considering the simple chain equivalent models
                respective to the situations in Figure~\ref{fig:res_orig}.
  }~\label{fig:res_equiv} 
  \end{center}
\end{figure*}

The hierarchical measurements of the structures considered above ---
namely the hierarchical number of nodes, hierarchical degree --- were
obtained and used to define the respective equivalent
models. Figure~\ref{fig:res_equiv} depicts the total number of spikes
along time obtained for the equivalent models with respect to the
situations shown in Figure~\ref{fig:res_orig}.  Strikingly, all plots
--- except possibly for the GG case --- show a clear transition of
spiking intensity near time step 80.  As expected, the WS and GG
networks yielded more gradual transitions.  However, in all cases the
network was completely activated after nearly 80 time steps,
undergoing saw oscillations.  Interestingly, the amplitude of such
oscillations varied considerably for each of the situations in
Figure~\ref{fig:res_equiv}, being more intense for the two knitted
models (i.e. PN and PA).  In addition, the height of the oscillating
plateaux tended to follow the intensity of the avalanches in
Figure~\ref{fig:res_orig}, being close to 50 for all models, except
for the PA model, which yielded a slightly lower avalanche transition
of nearly 40 units.

\begin{figure*}[htb] 
  \begin{center}
  \includegraphics[width=0.4\linewidth]{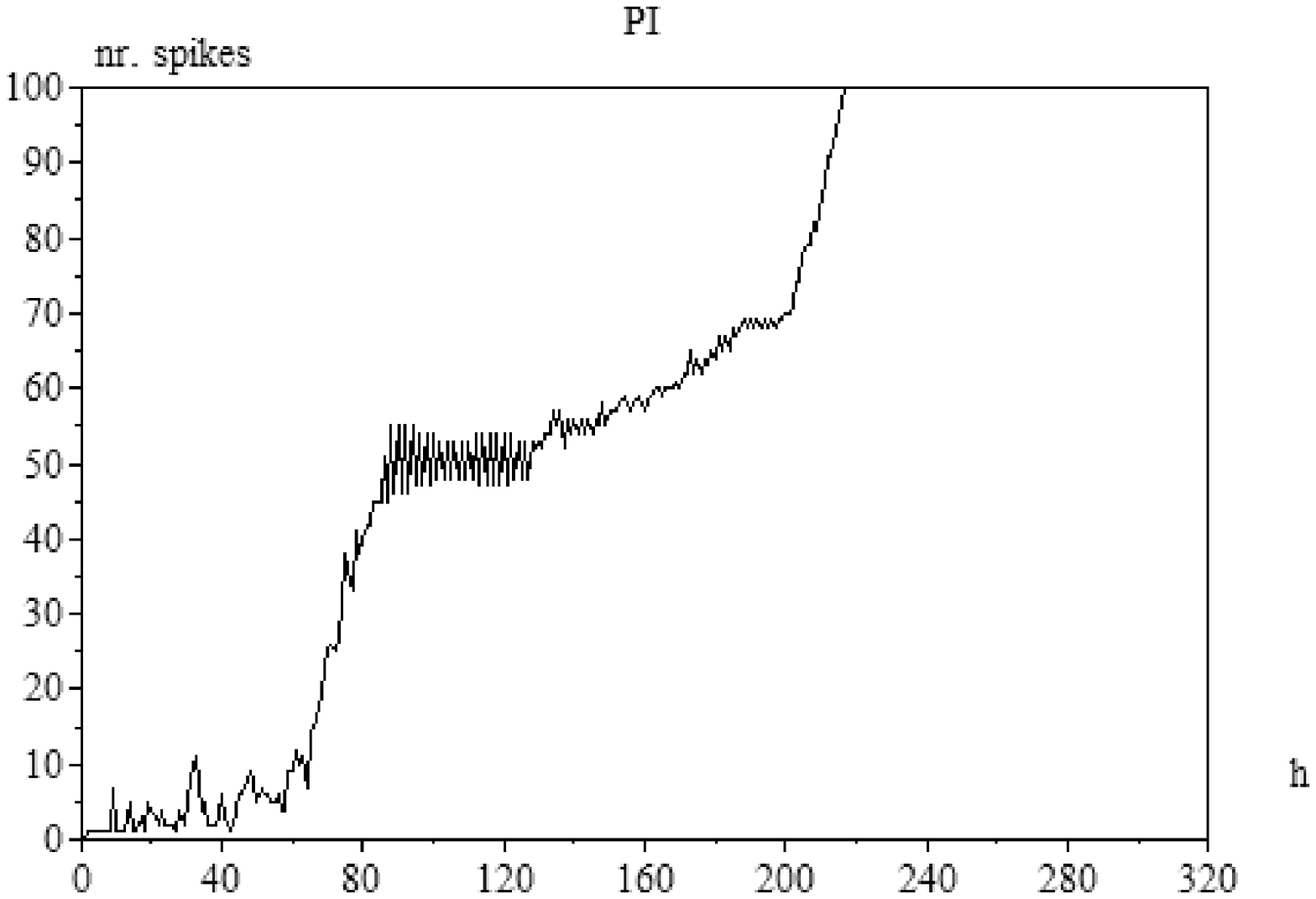}  \hspace{1cm}
  \includegraphics[width=0.4\linewidth]{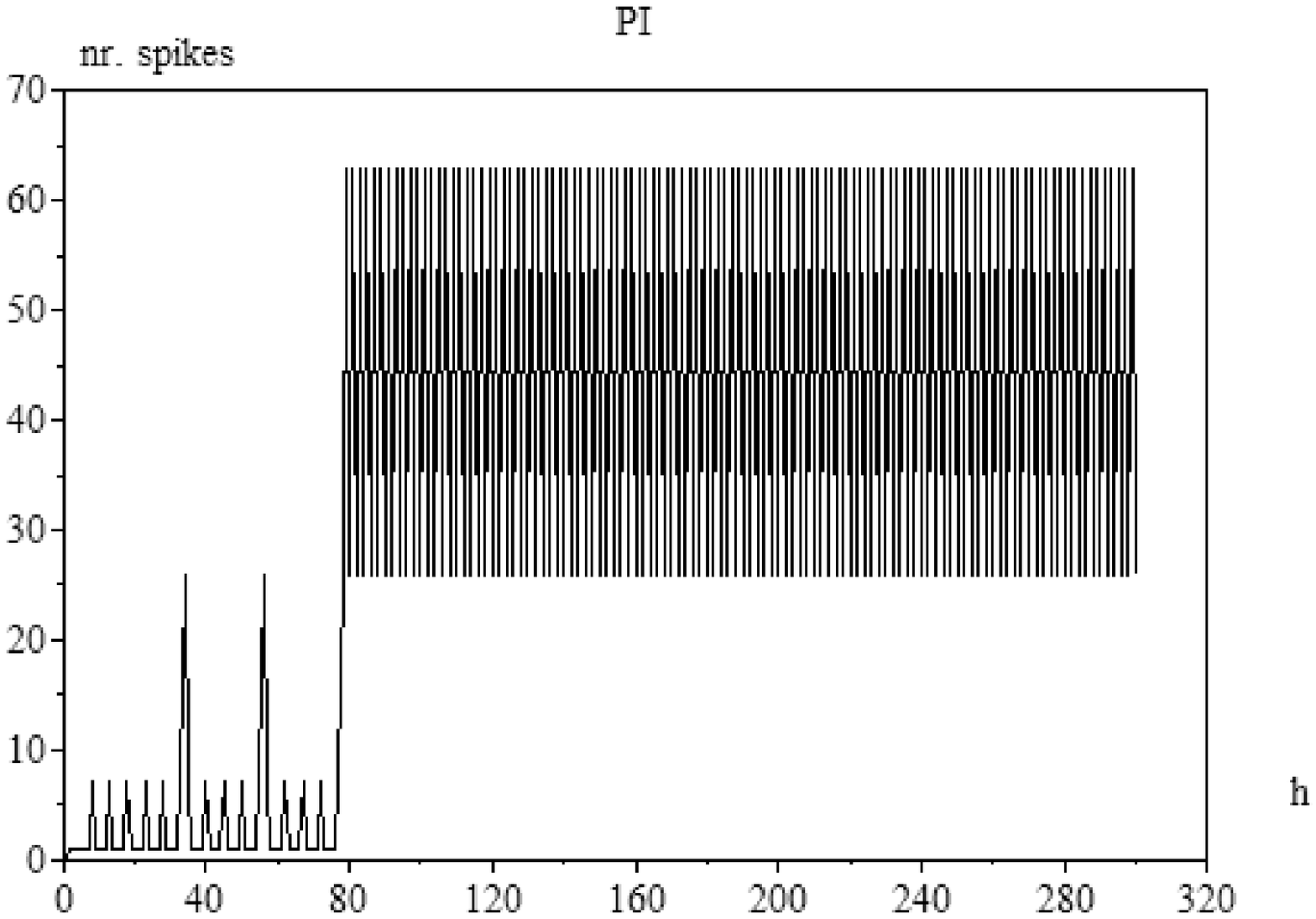}  \\
  (a)  \hspace{8cm}  (b)
  \caption{The total of spikes in terms of time obtained for the
                highly regular PI model considering the whole
                original network (a) and the repspective equivalent
                model (b). 
  }~\label{fig:res_PI} 
  \end{center}
\end{figure*}

Figure~\ref{fig:res_PI} shows the results obtained for the highly
regular PI model by using the whole original network (a) and the
respective equivalent model (b).  A pronounced avalanche is obtained,
followed by a relatively long plateau of stabilization of the overall
spiking rate.  A secondary avalanche can be identified at
approximately 210 steps.  Again, the equivalent model was capable of
identifying the timing and intensity of the first avalanche.

By being in complete agreement, as far as the main avalanche
properties are concerned, with the results obtained by considering the
whole original network (Fig.~\ref{fig:res_orig}), these results
support the fact that the intrinsic dynamics of avalanches, as well as
their respective times and intensities, are intrinsically captured in
the particularly simple respective equivalent models.  However, the
simple chain model failed to detect the secondary avalanches obtained
for the WS and PN models.  Remarkably, the intrinsic dynamics of the
first avalanches seem to be completely determined by the respective
topological measurements of hierarchical number of nodes and
hierarchical degrees.  This implies that the main avalanches, as well
as their features, can be reasonably predicted from the respective
hierarchical measurements.

\section{How universal are avalanches?}

The remarkable results obtained in the previous section, especially
the similar onset times for the avalanches obtained for all models
except GG, motivate future investigations.  In this section, we
consider the highly regular PI model in order to identify the
equivalent model parameters (i.e. hierarchical number of nodes and
hierarchical signatures) for a whole range of network sizes and
average degrees (recall that all nodes in a PI network have the same
degree).  

\begin{figure}[htb] 
  \begin{center}
  \includegraphics[width=0.8\linewidth]{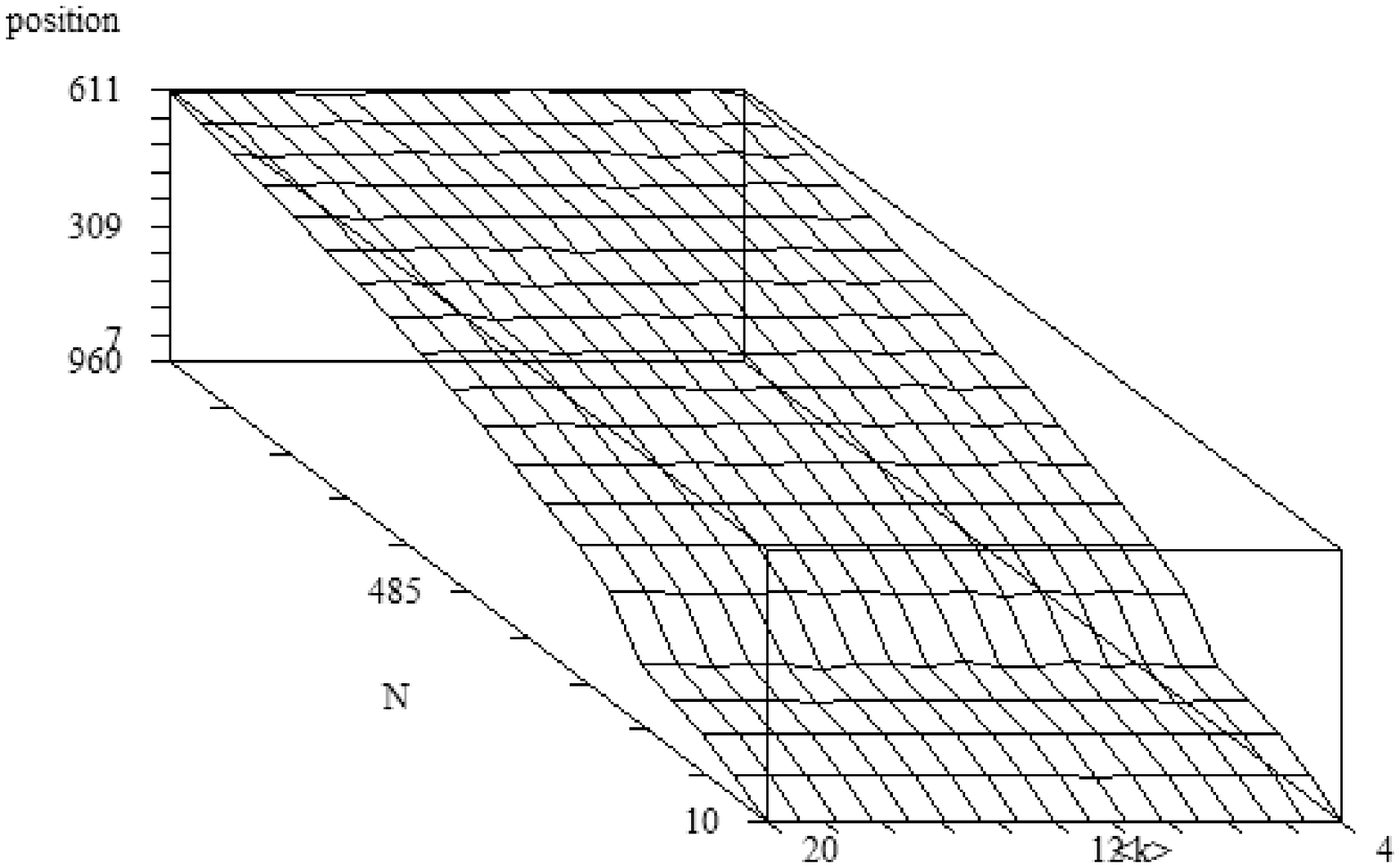}  \\
    (a) \\
  \includegraphics[width=0.8\linewidth]{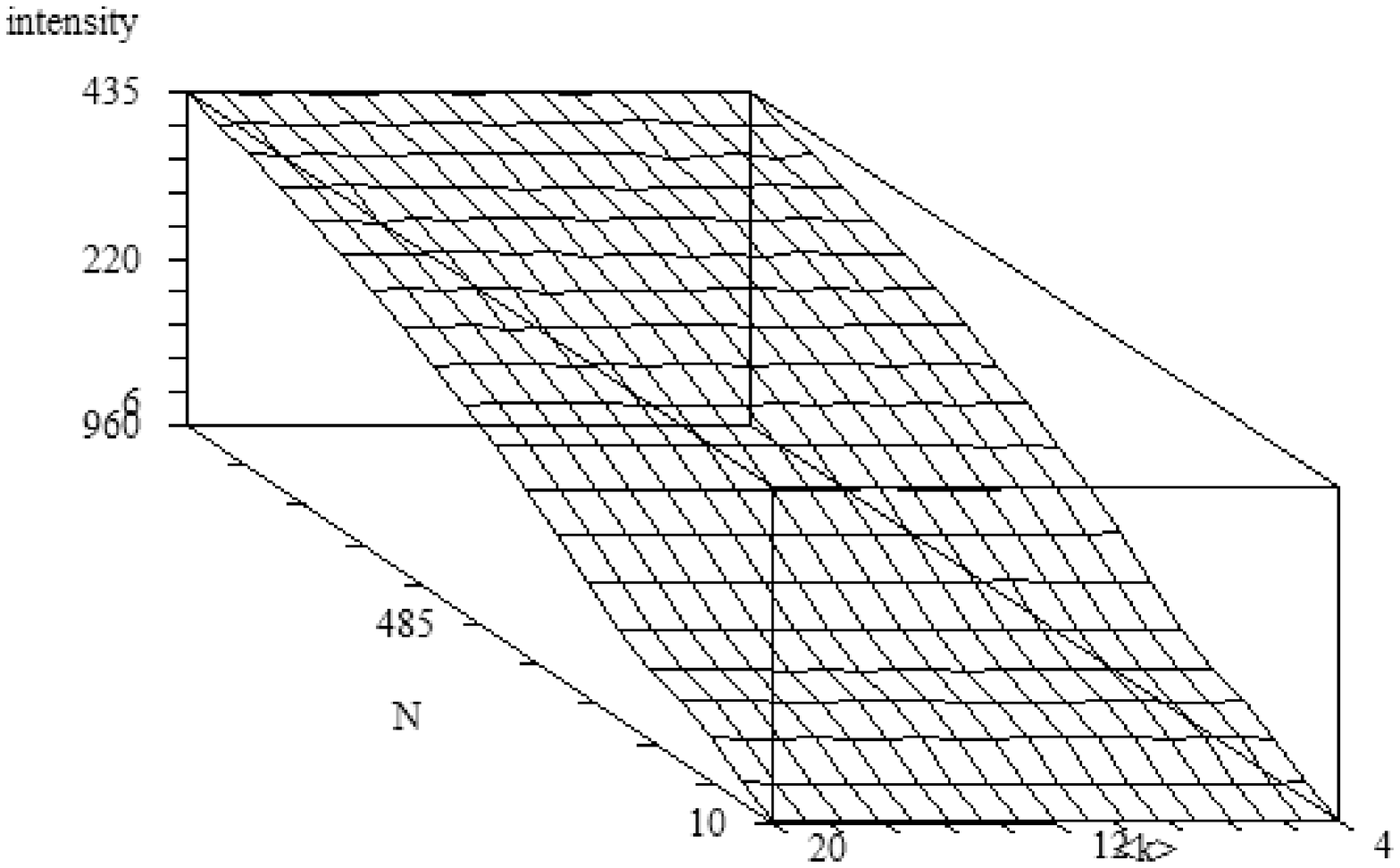}  \\
    (b) \\
  \caption{The position (a) and intensity (b) of the
                avalanches predicted for the PI networks
                of varying sizes ($N$) and average degrees 
                ($\left< k \right>$). 
  }~\label{fig:NN} 
  \end{center}
\end{figure}

Figure~\ref{fig:NN} shows the positions and intensities predicted for
the avalanches in PI complex neuronal networks in terms of their sizes
$N$ and average node degree $\left< k \right>$.  These results were
obtained by considering the whole networks (not the equivalent models)
and averaging over all nodes of the PI networks.  Interestingly, both
the positions and the intensities do not depend on the average degree.
This is compatible with the fact that only the \emph{ratios} between
the hierarchical degrees are taken into account in the equivalent
model.  The positions of the avalanches along time
(Fig.~\ref{fig:NN}a) increase monotonically, but not linearly, with
$N$.  The positions also undergo an abrupt increase near $N=100$.  The
intensities of avalanches (Fig.~\ref{fig:NN}b) exhibit a similar
behavior, also independent of the average degree, but exhibit a less
abrupt increase near $N=100$.

\section{A Real-World Example: \emph{C. elegans} Network}

Having proceeded all the way from the discussion of the avalanches in
integrate-and-fire complex neuronal networks to the respective
characterization in terms of the chain equivalent model, it is now
time to illustrate the potential of these concepts and approaches with
respect to a real-world network, namely the \emph{C. elegans}
network~\cite{Watts_Strogatz:1998}.  The largest component considered
in this article contained 246 nodes (the original matrix was
thresholded at 4).

\begin{figure}[htb] 
  \begin{center}
  \includegraphics[width=0.8\linewidth]{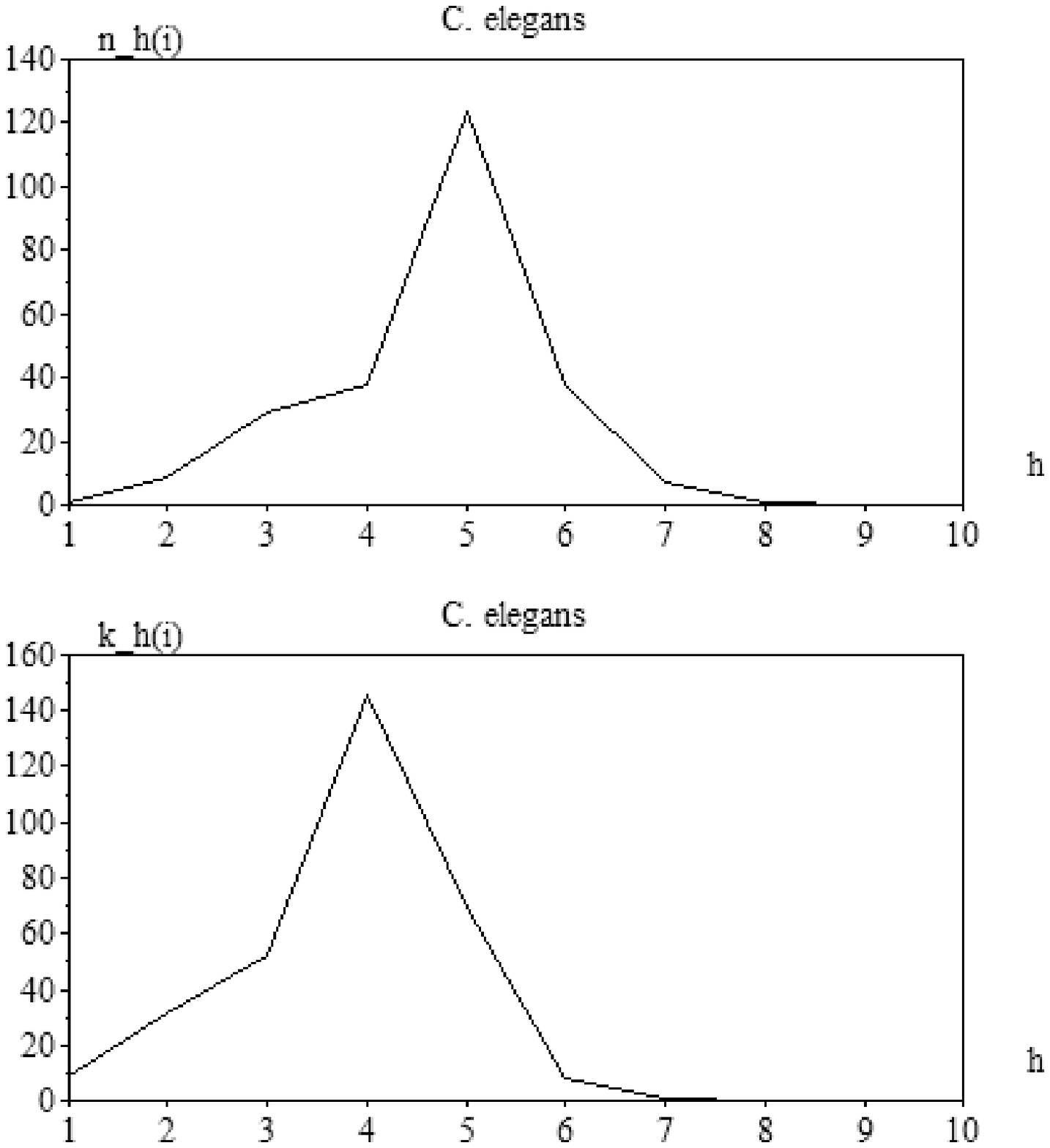}
  \caption{The hierarchical number of nodes and hierarchical degrees of
  the \emph{C. elegans} network considering node 50 as the reference.
  }~\label{fig:conc_celegans} \end{center}
\end{figure}

\begin{figure*}[htb] 
  \begin{center}
  \includegraphics[width=0.4\linewidth]{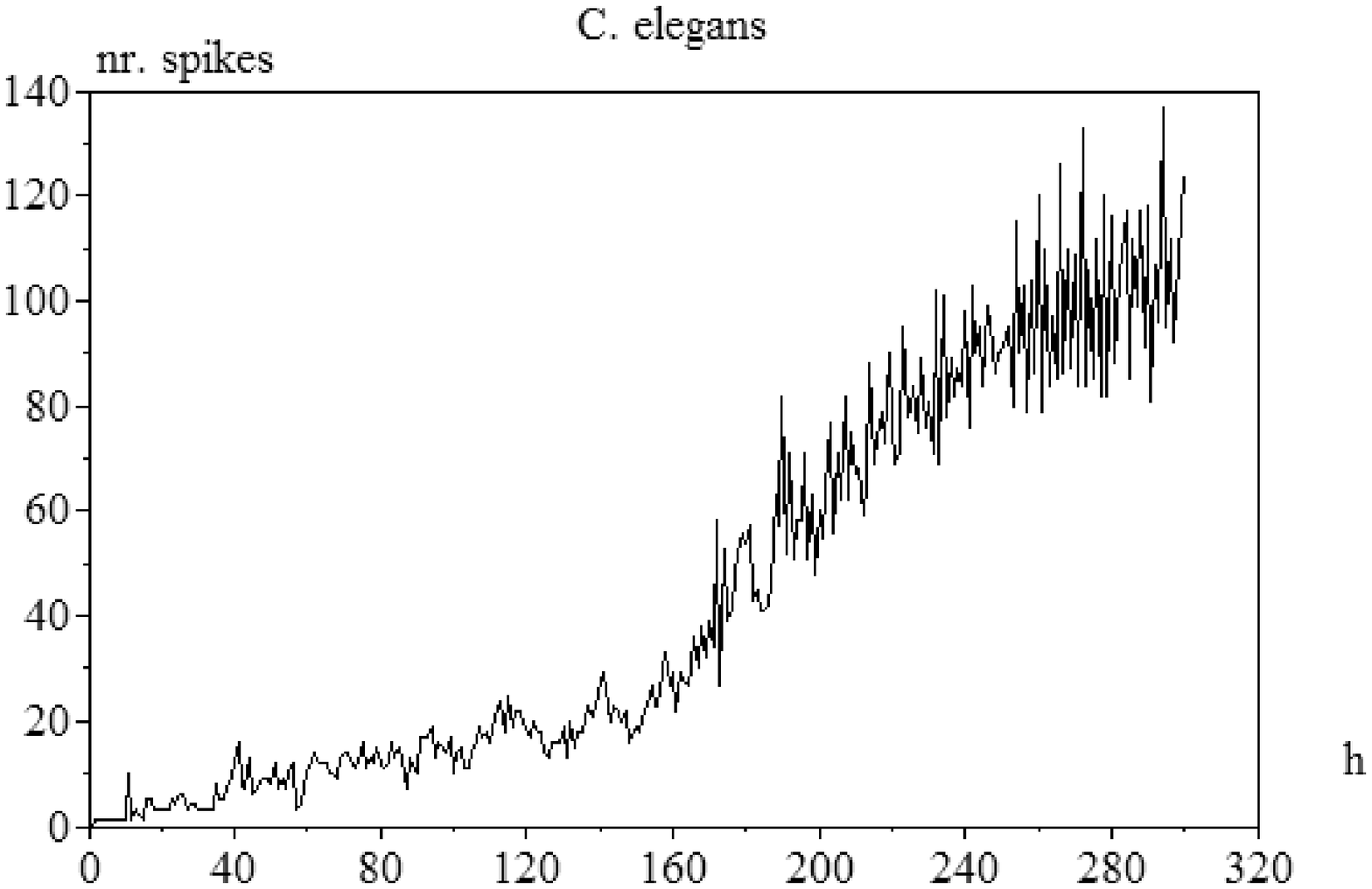}  \hspace{2cm}
  \includegraphics[width=0.4\linewidth]{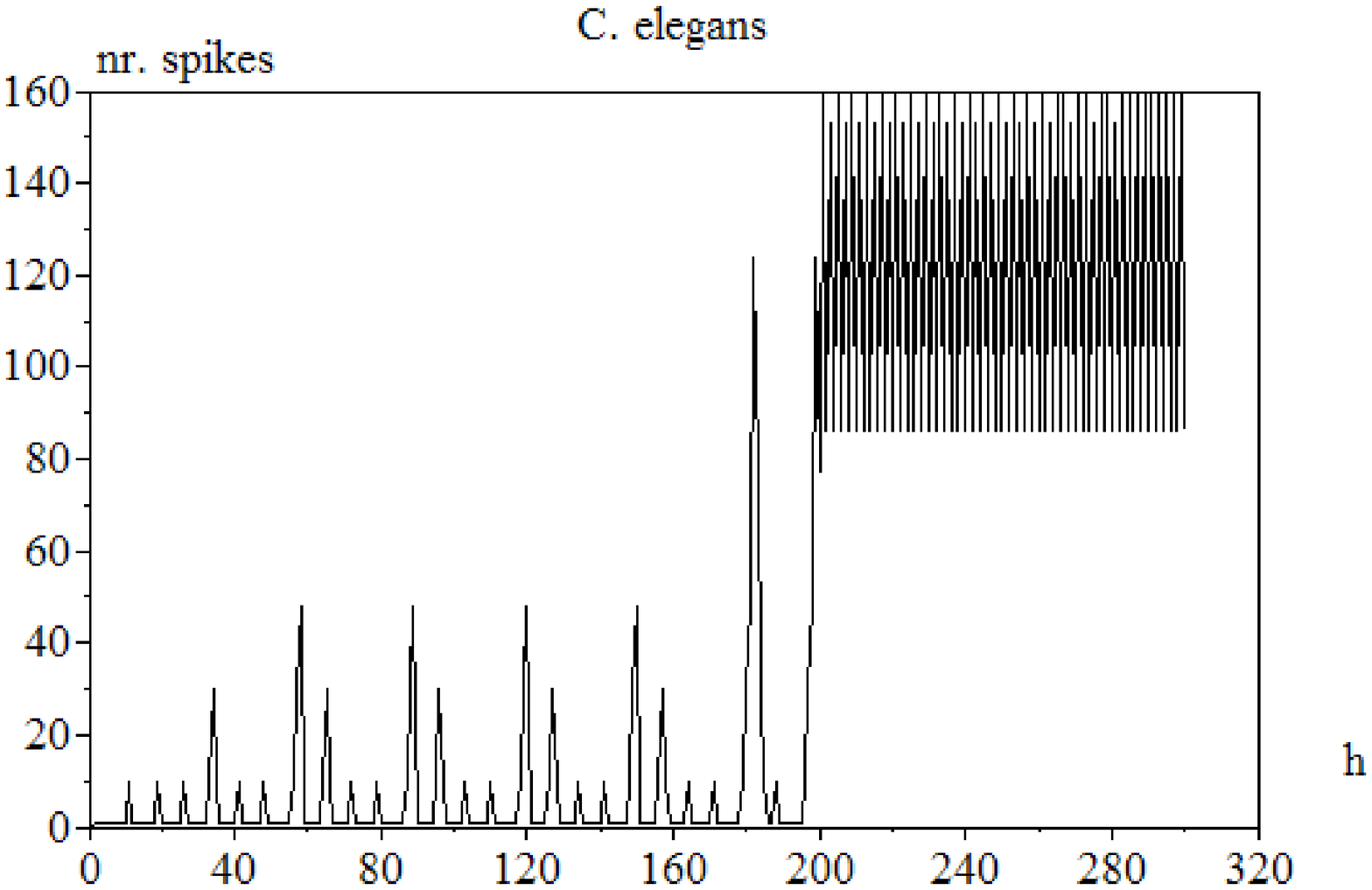}  \\
  (a) \hspace{8cm} (b)
  \caption{The total number of spikes obtained for the \emph{C. elegans}
                network with the activation source placed at node 50
                obtained by considering the whole network (a) and the
                respective equivalent model (b) defined by the hierarchical
                features in Figure~\ref{fig:conc_celegans}.
  }~\label{fig:celegans_avals} 
  \end{center}
\end{figure*}

Figure~\ref{fig:conc_celegans} shows the hierarchical number of nodes
$n_h(i)$ and hierarchical degrees $k_h(i)$ obtained for the
\emph{C. elegans} network with reference to node 50.  The maximum 
number of nodes per concentric level is obtained for level 5.  So, the
avalanche time is predicted as the sum of the nodes at this and the
previous levels, therefore corresponding to 200 steps.  The intensity
of the avalanche is estimated as being equal to the maximum number of
nodes per concentric level,
i.e. 123. Figure~\ref{fig:celegans_avals}(a) shows the total number of
spikes along time obtained considering the whole original structure
(a) and by using the equivalent model (b) defined by the respective
hierarchical measurements.  The avalanche is observed very close to
the predicted time, exhibiting intensity also very similar to the
respectively predicted value.

\begin{figure}[htb] 
  \begin{center} \includegraphics[width=0.8\linewidth]{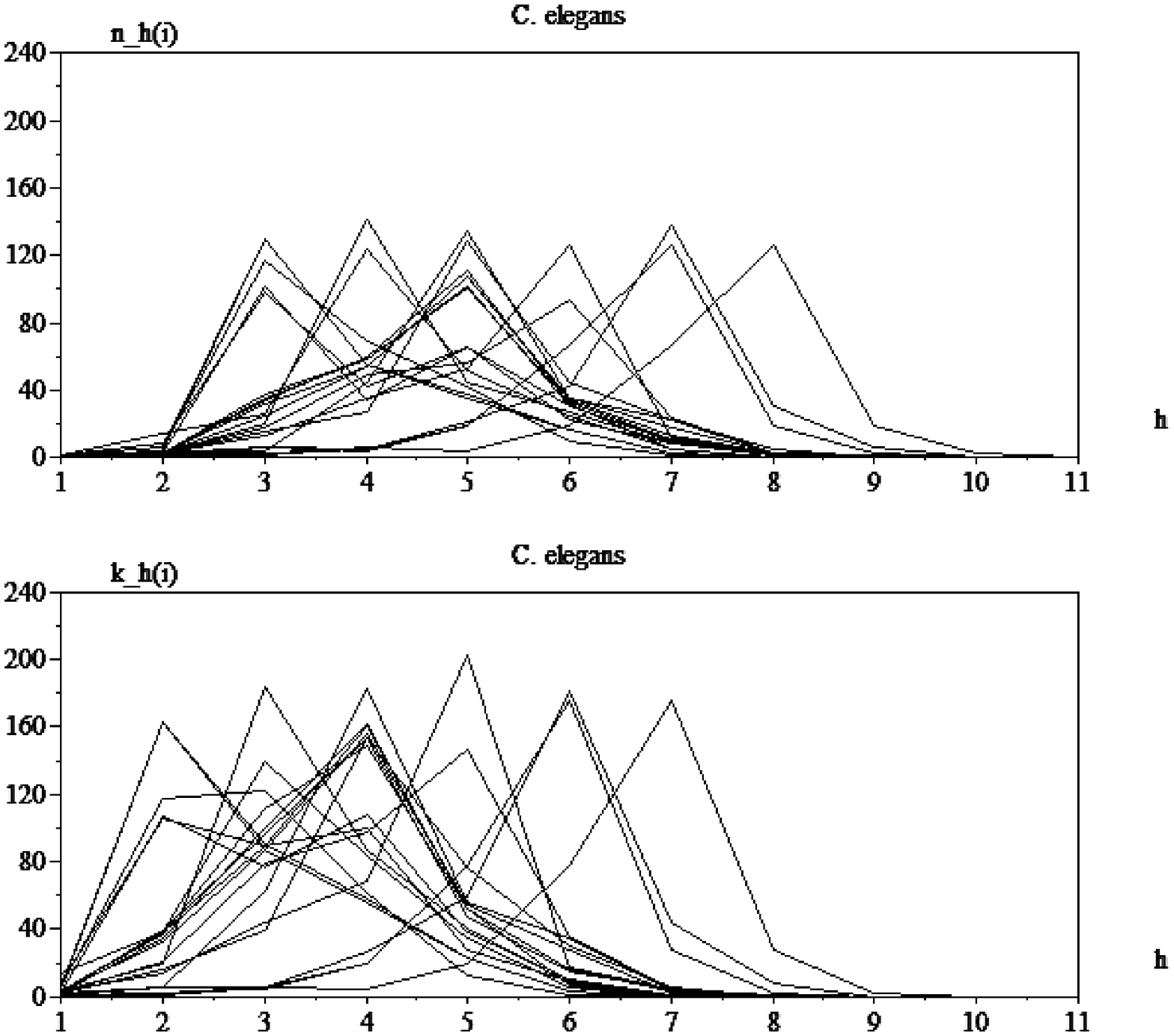}
  \caption{The hierarchical number of nodes and hierarchical degrees of
  the \emph{C. elegans} network considering node 50 as the reference.
  }~\label{fig:other_nodes} \end{center}
\end{figure}

Because of the close relationship between the concentric organization
and the transient integrate-and-fire dynamics, it is possible to
predict with accuracy the avalanche features from the hierarchical
number of nodes and hierarchical degrees of other nodes from the
\emph{C. elegans} network, without the need to perform the simulation 
of the respective dynamics.  Figure~\ref{fig:other_nodes} shows the
hierarchical number of nodes and hierarchical degrees obtained for 50
other nodes of the \emph{C. elegans} neuronal network.  It is clear
from such results that the placement of the activation at other nodes
will clearly lead to different avalanche intensities and positions
along time. This has been experimentally verified but is not reported
here for the sake of space.

\section{Concluding Remarks}

The integration of neuronal and complex networks provides an
especially exciting prospect for investigating the relationship
between structure and dynamics in complex systems.  Recently reported
results considering transient non-linear dynamics in
integrate-and-fire complex neuronal networks have yielded a series of
remarkable results, including the identification of avalanches of
activation and spikings occurring along time for several network
topologies~\cite{Costa_nrn:2008} and the transient confinement of
activation/spiking within the topological modules (communities) of
networks~\cite{Costa_begin:2008, Costa_activ:2008}.  The present work
had as its main objective to investigate the former of these effects.

After presenting the basic concepts, a series of developments were
reported with respect to the study of the first spiking avalanche in
complex neuronal networks.  The main contributions of the current
article are summarized as follows:

{\bf Introduction of the highly PI network model:} A new network
model, characterized by identical node degrees as well as other types
of structural regularities, has been introduced.  The PI model, which
is a version of the path-regular knitted
network~\cite{Costa_path:2007}, is obtained by implementing paths
involving all network nodes, without repetition of nodes or
\emph{edges}.  The PI model is particularly relevant for the type of
investigations developed in this article because it yields regular
degrees at all concentric levels.  Were not for the intra-ring links,
the PI model would lead to completely simultaneous spikings at each
concentric level.

{\bf Study of spiking dynamics for two extreme dual situations:} The
phenomenon of the spiking avalanches has been investigated and
characterized with respect to two extreme situations, namely a network
involving a hub (star connectivity) and a chain network.  Completely
different activations and spiking patterns are implied by each of
these two cases, corroborating the dual nature of star and path
connectivity~\cite{Costa_path:2007}.  More specifically, while the hub
network yields a maximally intense and simultaneous avalanche of
spikes and activation, the chain network implies gradual and smooth
propagation of the activation and spiking throughout the network.

{\bf Identification of the saw oscillations:} The unfolding of the
activation in integrate-and-fire complex neuronal networks tends to
exhibit oscilations.  This effect has been discussed and
characterized with respect to the chain networks.  In particular, it
has been shown that such an alternating dynamics corresponds to a
stable attractor of the dynamics.

{\bf Identification of secondary avalanches:} A secondary avalanche of
spikings has been identified for some network models, especially the
WS and PN models.  Though not investigated in more detail in the
present work, and not being captured by the equivalent model, such
phenomena correspond to another interesting feature of the
integrate-and-fire neuronal complex networks.

{\bf Brief review of hierarchical measurements of complex networks:}
While traditional measurements such as the degree and clustering
coefficient provide valuable information about only the most immediate
neighborhood around each node, the hierarchical (or concentric)
measurements allow a comprehensive characterization of the
connectivity surrounding each node in terms of multiple topological
scales.  The concentric organization of networks, as well as the
measurements of number of nodes by level and the hierarchical degree,
have been reviewed in an introductory and didactic fashion.  They have
been illustrated with respect to each of the 7 theoretical models of
complex networks adopted in this work.  The signatures obtained for
the hierarchical number of nodes, hierarchical degrees, and intra-ring
degrees are remarkably similar for all models except the geographical
and Watts-Strogatz structures.  Such a topological feature suggests
that any dynamics relying strongly on the concentric structures of the
networks will tend to exhibit similar properties in most of the
considered networks.

{\bf The importance of hierarchical measurements for dynamics:}
Because the neuronal activation is assumed to emanate from a single
node, the concentric organization of the network defined by this node
provides a natural way to look at the relationship between the
topology and spread of activation through the network.  Indeed,
because of the integration of neuronal activation along time
(facilitation) required for reaching the firing threshold, each of the
concentric levels acts as a barrier, confining the neuronal activation
at the previous levels until the activation becomes critical, implying
the almost simultaneous firing of the neurons in the lastly activated
level. Therefore, the main avalanche during the transient activation
of a complex neuronal network is determined by the concentric level
containing the largest number of nodes.  Because the
integrate-and-fire dynamics in complex neuronal networks tend to be
strongly defined by the concentric structure around the source node,
several of the features of the avalanches can be estimated from such
topological information.  It should be observed that such a
formulation can be readily extended to consider more than one node
acting as sources, as allowed by the generalized approach where
subgraphs are treated as equivalent nodes~\cite{Costa_EPJB:2005}.

{\bf Proposal of a chain equivalent model:} The close relationship
between the topological concentric features of the networks and the
respective spread of activation and spiking motivated the proposal of
a simple equivalent model, namely a chain network whose each node
corresponds to one of the concentric levels of the original network
and the weights are defined by the proportion of activations sent to
the previous and next levels.  Though such a proportion has been
determined in terms of the hierarchical degrees at the previous and
next levels, because of the assumed degree regularity, it can also be
obtained from the number of nodes at the previous, current and next
levels. In order to account for the different number of nodes at each
of the levels of the original network, the previously adopted
integrate-and-fire neuron model was augmented to incorporate synaptic
weights.  By reducing the incoming activation by the right proportion,
determined in terms of the ratio between the hierarchical degree (or
number of nodes between layers), the synaptic weights mimic the large
inertia for activation implied by the different number of nodes at
each level.  Though very simple, typically involving only a handful of
nodes, the equivalent model was shown to capture the intrinsic
topological features defining important features of the avalanches.
More specifically, their intensity can be reasonably estimated as
corresponding to the number of nodes at the concentric level with the
largest number of nodes.  The time when the main avalanche occurs can
also be predicted by adding the number of nodes from the first level
to the level containing the largest number of nodes.  The equivalent
model was shown to predict these parameters with reasonable accuracy
for all 7 models as well as for the \emph{C. elegans} network, except
the geographical model (which do not yield avalanches).  Remarkably,
the equivalent model seems to be useful even for Barab\'asi-Albert
networks, which are characterized by highly skewed degree
distributions.  However, the equivalent model does not take into
account either intense degree non-uniformities or the intra-ring
connections.  In practice, such features tend to undermine the
simultaneous activation of the neurons in each concentric level,
making the avalanches less definite.

{\bf Identification of universal features of avalanches:} Because most
of the considered networks, including the \emph{C. elegans} structure,
exhibit similar concentric organizations with respect to most of their
nodes, and because the integrate-and-fire dynamics is strongly defined
by the concentric properties of the network with respect to the source
node, it followed that most of the considered models exhibit similar
non-linear transient dynamics.  The reported systematic investigation
of the intensity and position of the avalanches in the PI model
considering several network sizes and average degrees showed that
these properties are completely independent of the average degree, but
strongly affected by the network size. More specifically, both the
intensity and position of the main avalanche increases monotonically,
but sub-linearly with the network size.  Because of the fact that most
of the considered complex networks tend to exhibit similar concentric
organization, such results obtained for the PI model are very likely
immediately extensible to most models, except the Watts-Strogatz and
geographical structures.  This is an important result, suggesting some
level of universality among networks with the same number of nodes.
It is possible that such a universality is related to the stochastic
nature of the considered theoretical models.  It would be particularly
interesting to investigate if such a universality is verified for
real-world networks as well as other theoretical models.

{\bf Investigation of avalanches in the \emph{C. elegans} neuronal
network:} The consideration of this important real-world network
allowed additional insights about the avalanches as well as further
corroborated the potential of the equivalent model for predicting
important non-linear dynamical features with basis on the topological
concentric properties.  It has been further corroborated preliminary
results~\cite{Costa_nrn:2008}, by 	considering the source of activation at
node 50, that this network also undergoes an avalanche of spiking at
nearly the time step 200.  Similar results were obtained for other
nodes, though markedly distinct avalanche features were obtained for
some specific nodes.  This result is particularly striking because,
among the 7 considered theoretical models, only the GG and WS models
were characterized by different concentric signatures produced by
different nodes.  This would suggest the \emph{C. elegans} network to
be similar to either of those models.  However, those models did not
exhibit the avalanches verified for the
\emph{C. elegans} network. Consequently, it seems that this network
exhibits some specific topological structure which is not captured by
any of the 7 considered theoretical models. Such a heterogeneity of
dynamics implied by different nodes suggests investigations aimed at
quantifying the influence of each node on the overall transient
dynamics.

{\bf Relating network structure and dynamics:} Ultimately, this work
has lied at the heart of the structure-dynamics paradigm, addressing
transient non-linear dynamics in integrate-and-fire complex neuronal
systems.  More specifically, it has focused the specific phenomenon of
the activation/spiking avalanches previously identified for complex
neuronal networks~\cite{Costa_nrn:2008}.  Because of the intrinsic way
in which the integrate-and-fire non-linearity acts on the dynamics
while depending on the topology, the avalanches were ultimately found
to be strongly defined by the hierarchical (or concentric)
organization of complex neuronal networks, allowing the development of
an extremely simple chain equivalent model as well as the prediction
of the main avalanche features from the concentric measurements.

The results and concepts reported in this work have paved the way to
several future developments, some of which are discussed below:

{\bf Additional Experiments:} While the results reported in this work
were computationally limited, it would be important to extend the
reported investigations to include larger networks as well as a larger
number of samples for each configuration to be experimentally
simulated and characterized.

{\bf Extension to other types of dynamics:} The close relationship
between the concentric topology and dynamics identified in this work
suggests that other types of dynamics --- including traditional
diffusion, self-avoiding random walks, as well as integrate-and-fire
with activation decays~\cite{Costa_activ:2008}, may also be strongly
related to the hierarchical organization of complex networks.

{\bf Real-world networks:} The interesting results obtained for the
\emph{C. elegans} network motivate the investigation of other real-world 
networks related to non-linear dynamics.  Several cases would be
particularly interesting to be examined, including communications,
production, transportation and computing networks.  Of special
interest would be biological networks, where the production of
specific molecules and structures depend on the integration of
received subparts.  

{\bf Identification of nodes leading to particular types of
avalanches:} The verified fact that different nodes in some networks
(e.g. WS, GG and \emph{C. elegans}) can produce distinct avalanches
motivates further research aimed at the quantification and
identification of nodes in a given network which are capable of
producing the most intense avalanche, or the earliest and latest
avalanches.  By taking into account the respective effects of each
node on the dynamics, such an investigation could provide insights
about the role of these nodes in the respective real-world systems.

{\bf Incorporation of intra-ring edges and consideration of skewed
degree distributions:} For simplicity's sake, the equivalent model
reported in this work did not take into account the intra-ring
connections or non-uniformities of degree.  It would be interesting to
develop augmented models incorporating these features and to verify
whether they would be able to capture secondary avalanches and lead to
more precise estimations.

{\bf Extension to integrate-and-fire complex neuronal networks with
limitation of the activation transfers:} In this work we assumed that
once a spike occurs, the internal accumulated activation is integrally
shared between the outgoing axons.  However, in biological neuronal
networks, the axon activation is known to be limited to similar spike
amplitudes.  Preliminary investigations indicated that most of the
phenomena, concepts and results reported in this work extend
immediately to integrate-and-fire models involving outward activation
limitation.  Nevertheless, small differences have been verified, such
as the fact that in this type of limited activation structures the
avalanches tend to occur as large peaks, not as the sigmoid
transitions illustrated in this work.  Therefore, it would be
interesting to repeat the investigation procedure reported here in
order to characterize in more detail the relationship between topology
and dynamics in complex neuronal networks with outward activation
limitation.

{\bf The equivalent model and topological communities:} In addition to
advancing the understanding of the avalanches, the concepts and
results reported in the current work also paved the way to future
investigations of another remarkable effect of the non-linear
integrate-and-fire complex neuronal networks, namely the transient
confinement of activation within the topological communities.  Indeed,
it would be relatively easy to extend the adopted formulations to
incorporate groups of more intensely connected nodes at each of the
concentric levels.  By doing so, it is possible to define equivalent
models which involve several branches initiating at the source.

{\bf Superedge analysis of the synchronization of the activations:} It
would be particularly interesting to apply the concept of superedges,
recently suggested~\cite{Costa_superedges:2008} as a particularly
systematic and comprehensive way to investigate the relationship
between topology and dynamics in complex systems, in order to obtain
additional insights about the avalanches phenomenon.  At the same
time, the superedges approach can be complemented by the concentric
approaches currently adopted. The superedges approach would be
particularly important for quantifying the loss of spiking
simultaneity implied by non-uniform degree and intra-ring connections.

{\bf Implications for neuroscience, real-world networks, and complex
networks research:} The reported concepts and results have several
potential implications for neuroscience and complex networks research.
In the former case, it would be interesting to seek for avalanches in
the activation of biological neuronal systems and relate such
effects to the respective network topology.  Interestingly, because of
the prediction potential of the equivalent model and its simplicity,
it becomes possible to perform such investigations even when the
specific details of the biological networks are known. In other words,
\emph{avalanches of neuronal activity can be related to the network
architecture by considering estimations of the number of nodes in each
neuronal concentric level}.  Similar investigations can also be
performed with respect to a number of real-world networks involving
non-linear dynamics, especially other biological networks such as gene
expression and protein interaction. The implications for complex
network research are also numerous.  First, we have shown that the
intrinsic topological features of networks can have fundamental
effects on the transient activation of non-linear complex systems,
motivating further related works. Second, it has been shown relatively
complex phenomena such as avalanches in the transient activation of
non-linear integrate-and-fire complex neuronal networks, can be
effectively modeled by simple equivalent networks. In addition, it
would be interesting to seek for other near universal properties of
the avalanches. Finally, it has been shown that the concentric
organization of complex networks, which has been relatively overlooked
in the literature, can hold the key for explaining important dynamical
properties of complex neuronal networks.

\begin{acknowledgments}
Luciano da F. Costa thanks CNPq (308231/03-1) and FAPESP (05/00587-5)
for sponsorship.
\end{acknowledgments}

\bibliography{theon}
\end{document}